\documentclass[english,12pt]{article}
\usepackage{geometry}
\usepackage{color}
\pdfoutput=1
\usepackage{float}
\usepackage{graphicx}
\usepackage[numbers]{natbib}
\graphicspath{{\figures}}
\usepackage{subcaption}
\usepackage{url}
\usepackage{mathtools}
\usepackage{setspace}
\usepackage{amsfonts}
\usepackage{amsmath}
\usepackage{bm}
\usepackage{bbm}
\usepackage[normalem]{ulem}
\makeatletter
\renewcommand*\env@matrix[1][\arraystretch]{%
  \edef\arraystretch{#1}%
  \hskip -\arraycolsep
  \let\@ifnextchar\new@ifnextchar
  \array{*\c@MaxMatrixCols c}}
\makeatother

\makeatletter

\floatstyle{ruled}
\newfloat{algorithm}{tbp}{loa}
\providecommand{\algorithmname}{Algorithm}
\floatname{algorithm}{\protect\algorithmname}

\newcommand{\bfx}{{\bf x}}
\newcommand{\blambda}{\boldsymbol{\lambda}}
\newcommand{\G}{{\bf g}}
\date{}
\usepackage{algorithm,algpseudocode}
\usepackage{cases}
\usepackage{cleveref}
\algnewcommand{\Initialize}[1]{%
  \State \textbf{Initialize:}
  \Statex \hspace*{\algorithmicindent}\parbox[t]{.8\linewidth}{\raggedright #1}
}
\usepackage{algorithm,algpseudocode}
\usepackage{tensor}
\usepackage[noblocks]{authblk}
\author[1]{Raymond H. Chan\thanks{Research supported by HKRGC Grants No. CUHK14306316,  HKRGC CRF Grant C1007-15G, HKRGC AoE Grant AoE/M-05/12,
CUHK DAG No. 4053211, and CUHK FIS Grant No. 1907303. }}
\author[1]{Kelvin K. Kan\thanks{Research supported by US Air Force Office of Scientific Research under grant FA9550-15-1-0286.}}
\author[2]{Mila Nikolova\thanks{Research supported by the French Research Agency (ANR)
under grant No ANR-14-CE27-001 (MIRIAM)
and by the Isaac Newton Institute for Mathematical Sciences for
support and hospitality during the programme Variational Methods and Effective Algorithms for Imaging and Vision,
EPSRC  grant  no  EP/K032208/1.}}
\author[3]{Robert J. Plemmons\thanks{Research supported by HKRGC Grant No. CUHK14306316 and US Air Force Office of Scientific Research under grant FA9550-15-1-0286.}}
\affil[1]{Department of Mathematics, The Chinese University of Hong Kong, Hong Kong}
\affil[2]{CMLA, ENS Cachan, CNRS, Universit\'e Paris-Saclay, 94235 Cachan, France}
\affil[3]{Department of Computer Science and Department
of Mathematics, Wake Forest University, Winston-Salem, NC 27106,
USA}
\makeatother

\usepackage{babel}
\begin{document}
\title{A two-stage method for spectral-spatial classification of hyperspectral images}
\maketitle

\begin{abstract}
This paper proposes a novel two-stage method for the classification of hyperspectral images. Pixel-wise classifiers, such as the classical support vector machine (SVM), consider spectral information only; therefore they would generate noisy classification results as spatial information is not utilized. Many existing methods, such as morphological profiles, superpixel segmentation, and composite kernels, exploit the spatial information too. In this paper, we propose a two-stage approach to incorporate the spatial information. In the first stage, an SVM is used to estimate the class probability for each pixel.
The resulting probability map for each class will be noisy. In the second stage, a variational denoising method is used to restore these noisy probability maps to get a good classification map. Our proposed method effectively utilizes both spectral and spatial information of the hyperspectral data sets. Experimental results on three widely used real hyperspectral data sets indicate that
our method is very competitive when compared with current state-of-the-art methods,
especially when the inter-class spectra are similar or the percentage of the training pixels
is high.
\end{abstract}

\section{Introduction}
Remotely-sensed hyperspectral images (HSI) are images taken from airplanes or satellites
that record a wide range of electromagnetic spectrum, typically more than 100 spectral bands from visible to near-infrared wavelengths. Since different materials reflect different spectral signatures, one can identify the materials at each pixel of the image by examining its
spectral signatures. HSI is used in many applications, including agriculture \cite{Patel2001,Datt2003}, disaster relief \cite{Trierscheid2008,Eismann2009}, food safety \cite{Lu1999,Gowen2007}, military \cite{Manolakis2002,Stein2002} and    mineralogy \cite{Horig2001}.

One of the most important problems in hyperspectral data exploitation is HSI classification.
It has been an active research topic in past decades \cite{Mountrakis2011,Fauvel2013}. The pixels in the
hyperspectral image are labeled manually by experts based on careful review of the spectral signatures and investigation of the scene. Given these ground-truth labels (also called ``training pixels"), the objective of HSI classification is to assign labels to
part or all of the remaining pixels (the ``testing pixels") based on their spectral signatures
and their locations.

Numerous methods have been developed for HSI classification. Among these,
machine learning is a well-studied approach. It includes multinomial logistic regression \cite{Li2010,Li2012,Li2013}, artificial neural networks \cite{Benediktsson2005,Yue2015,Makantasis2015,Morchhale2016,Pan2017}, and support vector machines (SVMs) \cite{Boser1992,Cortes1995,Scholkopf2000}. Since our method is partly based on SVMs, we will
discuss it in more details here. The original SVM classification method  \cite{Melgani2004,Camps2005} performs pixel-wise classification that utilizes spectral information but not spatial dependencies. Numerous spectral-spatial SVM classification methods have been introduced since then. They show better performances when compared to the pixel-wise SVM classifiers. Here we report some of them.

SVMs with composite kernels \cite{Camps2006} use composite kernels that are weighted summations of spectral kernels and spatial kernels. The spatial information is extracted by taking the average of the spectra in a fixed window around each pixel. To further utilize the spatial information, the method in \cite{Fang2015G} first applies superpixel segmentation to break the hyperspectral image into small regions with flexible shapes and sizes. Then it extracts the spatial information based on the segmentation and finally performs the classification using SVMs with multiple kernels.
In \cite{Tarabalka2009}, a pixel-wise SVM classification is first used to produce classification maps, then a partitional clustering is applied to obtain a segmentation of the hyperspectral image. Then a majority vote scheme is used in each cluster and finally a filter is applied to denoise the result. The method in \cite{Kang2014} first produces pixel-wise classification maps using SVMs and then applies edge-preserving filtering to the classification maps. In addition to these methods, techniques based on Markov random fields \cite{tarabalka2010}, segmentation \cite{Tarabalka2009,Ghamisi2014,Fang2015G,Liu2017} and morphological profiles \cite{Fauvel2008,Liu2017} have also been incorporated into SVMs to exploit the spatial information.

Besides machine learning approaches, another powerful approach is sparse representation \cite{Bruckstein2009}. It is based on the observation that spectral signatures within the same class usually lie in a low-dimensional subspace; therefore test data can be represented by a few atoms in a training dictionary. A joint sparse representation method is
introduced in \cite{Chen2011} to make use of the spatial homogeneity of neighboring pixels. In particular, each test pixel and its neighboring pixels inside a fixed window are jointly sparsely represented. In \cite{Chen2013}, a kernel-based sparse algorithm is proposed
 which incorporates the kernel functions into the joint sparse representation method.
It uses a fixed size local region to extract the spatial information. Approaches with more flexible local regions were proposed in \cite{Fang2014} and \cite{Fang2015}. They incorporate a multiscale scheme and superpixel segmentation into the joint sparse representation method respectively. Multiple-feature-based adaptive sparse representation was proposed in \cite{Fang2017}. It first extracts various spectral and spatial features and then the adaptive sparse representations of the features are computed. The method in \cite{Li2016} first estimates the pixel-wise class probabilities using SVMs, then applies sparse representation to obtain superpixel-wise class probabilities in which spatial information is utilized and the final result is obtained by combining both probabilities.

A pixel-wise classifier (such as SVM), which considers only spectral information, generates  results with decent accuracy but would appear noisy as spatial information is not used, see \cite{Melgani2004} and also \Cref{{example_noisy_pixelSVM}}. The noise can be restored by image denoising techniques that incorporate the spatial information. Image denoising is a well-studied subject and numerous effective denoising methods  have been introduced \cite{Chan2000,Nikolova2004,Chan2005,Hintermuller2006,Bredies2010}. In this paper, we propose a simple but effective two-stage classification method inspired by our two-stage method for impulse noise removal \cite{Chan2005}. In the first stage, we apply a pixel-wise SVM method that exploits the
spectral information to estimate a pixel-wise probability map for each class. In the second stage, we apply a convex denoising model to exploit the spatial information
so as to obtain a smooth classification result. In the second stage, the training pixels are kept fixed as their ground-truth labels are already given. In this sense, this stage is exactly the
same at the second stage in our impulse noise removal method in \cite{Chan2005}.

\begin{figure}[h!]
        \centering
            \includegraphics[scale=1.3]{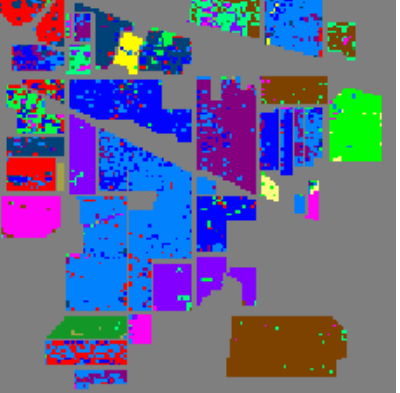}
            \caption[]%
            {{\small An example of classification result using pixel-wise SVM classifier\label{example_noisy_pixelSVM}}}
        \end{figure}

Our method utilizes only spectral information in the first stage and spatial information in the second stage. Experiments show that our method generates very competitive accuracy compared to the state-of-the-art methods on real HSI data sets,
especially when the inter-class spectra are similar or
the percentage of training pixels is high. This is because our method
can effectively exploit the spatial information
even when the other methods cannot distinguish the spectra. Moreover,
our method has small number of parameters
and shorter computational time than the state-of-the-art
methods.

This paper is organized as follows. In \Cref{sec:review} the support vector machine and variational denoising methods are reviewed. In \Cref{sec:proposed_method} our proposed two-stage classification method is presented. In \Cref{sec:results} experimental results are presented to illustrate the effectiveness of our method. \Cref{sec:conclusion} concludes the paper.

\section{Support Vector Machines and Denoising Methods}\label{sec:review}

\subsection{Review of $\nu$-Support Vector Classifiers}
Support vector machines (SVMs) has been used successfully in
pattern recognition \cite{Pontil1998}, object detection \cite{El2002,Osuna1997},
and financial time series forecasting \cite{Tay2001,kim2003} etc.
It also has superior performance in hyperspectral classification especially when the dimensionality of data is high and the number of training
data is limited \cite{Melgani2004,Camps2005}.
In this subsection, we review  the $\nu$-support vector classifier ($\nu$-SVC) \cite{Scholkopf2000}
which will be used in the first stage of our method.

Consider for simplicity a supervised binary classification problem.
We are given $m$ training data $\{\mathbf{x}_{i}\}_{i=1}^m$ in $\mathbb{R}^{d}$,
and each data is associated with a binary label $y_i \in \{-1,+1\}$ for
$i=1,2,...,m$. In the training phase of SVM, one aims to find a hyperplane
to separate the two classes of labels and maximize the distance between the
hyperplane and the closest training data, which is called the support
vector. In the kernel SVM, the data is mapped to a higher dimensional
feature space by a feature map $\phi:\mathbb{R}^d\rightarrow\mathbb{R}^h$
in order to improve the separability between the two classes.

The $\nu$-SVC is an advanced support vector classifier which enables the user to specify the maximum training error before the training phase. Its formulation is given as follows:
\begin{equation}\label{nu_svm}
\begin{cases}
\hfil \underset{\mathbf{w},b,\xi,\rho}{\text{min }}\, \frac{1}{2}||\mathbf{w}||_2^2 - \nu \rho + \frac{1}{N} \sum\limits_{i=1}^{m} \xi_i \\
\text{subject to: } y_i(\mathbf{w}\cdot \phi(\mathbf{x}_{i})+b)\geq \rho-\xi_{i},\; i=1,2,\ldots,m, \\
\hfil \xi_i \geq 0, \; i=1,2,\ldots,m, \\
\hfil \eta \geq 0,
\end{cases}
\end{equation}
where $\mathbf{w} \in \mathbb{R}^{h}$ and $b\in \mathbb{R}$ are
the normal vector and the bias of the hyperplane respectively,
$\xi_i$'s are the slack variables which allow training errors, and
 $\rho/||\mathbf{w}||_2$ is the distance between the hyperplane and the support vector. The parameter $\nu\in(0,1]$ can be shown to be an upper bound on the fraction of training errors \cite{Scholkopf2000}.

The optimization problem (\ref{nu_svm}) can be solved through its Lagrangian dual:
\begin{equation}\label{dual_nu_svm}
\begin{cases}
\hfil \underset{\mathbf{\alpha}}{\text{max }}\, -\frac{1}{2}\sum\limits_{i,j=1}^{m} \alpha_{i}\alpha_{j}y_{i}y_{j}K(\mathbf{x}_i,\mathbf{x}_j) \\
\text{subject to: } 0\leq \alpha_i \leq \frac{1}{N},\; i=1,2,\ldots,m,\\
\hfil \sum\limits_{i=1}^{m}\alpha_i y_i=0, \\
\hfil \sum\limits_{i=1}^{m} \alpha_i \geq \nu.
\end{cases}
\end{equation}
Its optimal Lagrange multipliers can be calculated using quadratic programming methods \cite{Vapnik1998}. After obtaining them, the parameters of the optimal hyperplane can be represented by the Lagrange multipliers and the training data. The decision function for a test pixel $\mathbf{x}$ is given by:
\begin{equation}\label{eq:SVM_decision}
g(\mathbf{x})=\text{sgn}(f(\mathbf{x})),\;\text{where }f(\mathbf{x})=
\sum\limits_{i=1}^{m} \alpha_i y_i K(\mathbf{x}_i,\mathbf{x})+b.
\end{equation}

Mercer's Theorem \cite[p.~423-424]{Vapnik1998} states that a symmetric function $K$ can be represented as an inner product of some feature maps $\phi$, i.e. $K(\mathbf{x},\mathbf{y})=\phi(\mathbf{x})\cdot \phi(\mathbf{y})$ for all $\mathbf{x},\mathbf{y}$, if and only if $K$ is positive semi-definite. In that case, the feature map $\phi$ need not be known in order to perform the training and classification, but only the kernel function $K$ is required. Examples of $K$  satisfying the condition in Mercer's Theorem include:
$K(\mathbf{x}_i,\mathbf{x}_j) = \text{exp}(-||\mathbf{x}_i-\mathbf{x}_j||^2/(2\sigma^2))$
and
$K(\mathbf{x}_i,\mathbf{x}_j) =(\mathbf{x}_i \cdot \mathbf{x}_j)^p$.

\subsection{Review of Denoising Methods}
Let $\Omega=\{1,...,N_1\}\times\{1,...,N_2\}$ be
the index set of pixel locations of an image, ${\bf v}$ is the noisy image and ${\bf u}$ is the restored image.
One famous approach for image denoising is the total variation (TV) method. It involves an optimization model with a TV regularization term which corresponds to the function $\|\nabla \cdot \|_1$. However, it is known that it reproduces images with staircase effect, i.e. with piecewise constant regions.  Here, we
introduce two approaches to improve it and they  are related to our proposed method.

The first approach is to add a higher-order term, see, \textit{e.g.},   \cite{Mumford1994,Chan2000,Shen2003,Hintermuller2006,Bredies2010}.
In \cite{Hintermuller2006}, the authors considered minimizing
\begin{equation}\label{convex_MF}
H(\mathbf{u})=\frac{1}{2}||\mathbf{v}-\mathbf{u}||_2^2+\alpha_1||\nabla \mathbf{u}||_1+\frac{\alpha_2}{2}||\nabla \mathbf{u}||_2^2.
\end{equation}
Here the first term is the $\ell_2$  data-fitting term that caters for Gaussian noise.
The second term is the TV term while the third term is the extra higher order term added
to introduce smoothness to the restored image $\mathbf{u}$.
By setting the parameters $\{\alpha_i\}_{i=1}^2$ appropriately, one can control the trade off between a piece-wise constant and a piece-wise smooth $\mathbf{u}$.
In \cite{Cai2013,Chan2014,Cai2017}, the authors derived the same
minimizational function (\ref{convex_MF}) as a convex and smooth
approximation of the Mumford-Shad model for segmentation. They applied it successfully for segmenting greyscale and color images corrupted by different noise (Gaussian, Poisson, Gamma), information loss and/or blur successfully.

The second approach is to smooth the TV function $\|\nabla \cdot\|_1$.
In \cite{Chan2005},  a two-stage method is proposed to
restore an image ${\bf v}$ corrupted by impulse noise.
In the first stage an impulse noise detector called Adaptive
Median Filter \cite{Hwang1995} is used to detect the locations of possible noisy pixels. Then in the second stage, it restores the noisy pixels
while keeping the non-noisy pixels unchanged
by minimizing:
\begin{align}\label{eq:l1_approx_restrict}
\begin{split}
F(\mathbf{u})& = ||\mathbf{v}-\mathbf{u}||_1+\frac{\beta}{2} \| \nabla \mathbf{u}\|^{\alpha},\\
\text{s.t. } & \mathbf{u}|_{\Upsilon}=\mathbf{v}|_{\Upsilon},
\end{split}
\end{align}
where $\Upsilon$ is the set of non-noisy pixels, $\mathbf{u}|_{\Upsilon}=(u_i)_{i\in \Upsilon}$, and $1 < \alpha \le 2$. This 2-stage method is the first method that can successfully restore images
corrupted with extremely high level of impulse noise (e.g. 90\%).

Our proposed method is inspired by this two-stage method.
In the first stage we use the spectral classifer $\nu$-SVC to generate a pixel-wise probability map for each class. Then in the second stage, we
use a combination of (\ref{convex_MF}) and (\ref{eq:l1_approx_restrict}) to restore the mis-classified pixels, subject to the constraint
that the training pixels are kept unchanged since their ground-truth labels are already
given.

\section{Our Two-stage Classification Method}\label{sec:proposed_method}
SVMs yield decent classification accuracy \cite{Melgani2004}  but their results can be noisy (see
Figure \ref{example_noisy_pixelSVM}) since only spectral information is
used. We therefore propose to use a denoising scheme to incorporate
the spatial information into the classification.  Our method first estimate the pixel-wise probability map for each class using SVMs. Then the spatial positions of the
training data are used in the denoising scheme to effectively remove the noise in the map.

\subsection{First Stage: Pixel-wise Probability Map Estimation}
\subsubsection{SVM Classifier}

HSI classification is a multi-class classification but the SVM is a binary classifier.
To extend SVM to multi-class, we use the One-Against-One (OAO) strategy \cite{Hsu2002} where $[c(c-1)/2]$ SVMs are built to classify every possible pair of classes.
Here  $c$ is the number of classes. In this paper, we choose the SVM method $\nu$-SVC \cite{Scholkopf2000} with OAO strategy
for the HSI multiclass classification in our first stage. We remark that one can use
other SVMs or multiclass strategy such as the One-Against-All strategy in \cite{Hsu2002} instead. Moreover, the  basis function kernel (RBF kernel) is used as the kernel function in our SVM method. The RBF kernel is defined as:

\begin{equation}\label{RBFkernel}
K(\mathbf{x}_i,\mathbf{x}_j) = \text{exp}\Big(-\frac{||\mathbf{x}_i-\mathbf{x}_j||^2}{2\sigma^2}\Big).
\end{equation}

\subsubsection{Probability Estimation of SVM Outputs}

Given a testing pixel $\bfx$ and a SVM classifier with
decision function $f(\bfx)$ in (\ref{eq:SVM_decision}), we
can label $\bfx$ with a class according to
the sign of $f(\bfx)$, see \cite{Cortes1995}.  Under the OAO strategy,
there are $[c (c-1)]/2$ such pairwise functions $f_{i,j}$, $1\le i,j \le c$, $i\neq j$. We use them to estimate the probability ${p_i}$ that $\mathbf{x}$ is in the $i$-th class. The idea is given in \cite{SVM_bin_prob,Wu2004}.
We first estimate the pairwise class probability ${\rm Prob}(y=i \ | \ y=i \text{ or } y=j)$ by computing
\begin{equation}
r_{i,j}=\frac{1}{1+e^{\rho f_{i,j}(\mathbf{x})+\tau}},
\end{equation}
where $\rho$ and $\tau$ are computed by minimizing a negative log likelihood problem over all the training pixels \cite{SVM_bin_prob}.

Then the probability vector $\mathbf{p}=[p_1,p_2,...,p_c]^T$ of the testing
pixel ${\bf x}$ is estimated by solving:
\begin{align}
& \underset{\mathbf{p}}{\text{min}}\; \frac{1}{2}\sum_{i=1}^{c}\sum_{j\neq i}(r_{j,i}p_i-r_{i,j}p_j)^2, \nonumber \\
& \text{s.t. }p_i \geq 0, \forall i,\; \sum_{i=1}^{c}p_i=1. \label{eq:ori_coupling}
\end{align}
Its optimal solution can be obtained by solving  the following simple linear system, see \cite{Wu2004}:
\begin{equation}\label{eq:lin_coupling}
\begin{bmatrix}
Q & \mathbf{e} \\
\mathbf{e}^T & {0}
\end{bmatrix}
\begin{bmatrix}
\mathbf{p} \\ {b}
\end{bmatrix}
=
\begin{bmatrix}
\mathbf{0} \\
{1}
\end{bmatrix},
\end{equation}
where
\begin{align*}
Q_{ij} = \begin{cases}
\sum\limits_{s\neq i} r_{s,i}^2 \; &\text{if } i=j, \\
-r_{j,i}r_{i,j} & \text{if } i\neq j,
\end{cases}
\end{align*}
${b}$ is the Lagrange multiplier of the equality constraint in (\ref{eq:ori_coupling}), $\mathbf{e}$ is the $c$-vector of all ones, and $\mathbf{0}$ is the $c$-vector of all zeros. In our tests, the probability vectors $\bf{p}(\bfx)$ for all testing pixels $\bfx$ are computed by this method using the toolbox of LIBSVM library \cite{libsvm}.

We finish Stage 1 by forming the 3D tensor $\mathcal{V}$
where $\mathcal{V}_{i,j,k}$ gives the probability that pixel $(i,j)$
is in class $k$. More specifically, if pixel $(i,j)$ is a testing pixel, then  $\mathcal{V}_{i,j,:}=\mathbf{p}(\bfx_{i,j})$; if pixel $(i,j)$ is a training pixel belonging to the $c$-th class, then $\mathcal{V}_{i,j,c}=1$ and $\mathcal{V}_{i,j,k}=0$ for all other
$k$'s.

\subsection{Second Stage: Restoring the Pixel-wise Probability Map}

Given the probability tensor $\mathcal{V}$ obtained in Stage 1, one can obtain
an HSI classification by taking the maximum probability for each pixel \cite{Kang2014}. However,
the result will appear noisy as no spatial information is taken into account.
The goal of our second stage is to incorporate the spatial information
into $\mathcal{V}$ by a variational denoising method that keeps the value of the training pixels unchanged during the optimization, as their ground-truth labels are given a priori.

Let $\textbf{v}_{k}:=\mathcal{V}_{:,:,k}$ be the ``noisy" probability map of the $k$-th class, where $k=1,...,c$. We restore them by minimizing:
\begin{align}
\label{eqn:l2l1anisoll2_eq}
\begin{split}
\underset{\mathbf{u}}{\text{min }}&\frac{1}{2}||\mathbf{u}-\mathbf{v}_k||_2^2+\beta_{1}||\nabla \mathbf{u}||_1+\frac{\beta_{2}}{2}||\nabla\mathbf{u}||_2^2,\\
\text{s.t. }& \mathbf{u}|_{\Upsilon}=\mathbf{v}_k|_{\Upsilon},
\end{split}
\end{align}
where $\beta_1$, $\beta_2$ are regularization parameters and ${\Upsilon}$ is the set of training pixels. We choose this minimization functional because it gives
superb performance in denoising and segmenting various types of images,
see \cite{Hintermuller2006,Cai2013,Chan2014,Cai2017}. The  higher-order
$||\nabla\mathbf{u}||_2^2$ term encourages smoothness of the solution and can improve the classification accuracy, see \Cref{sec:effect_higher_order}. In our tests, we use anisotropic TV \cite{Zhao2013} and periodic boundary condition for the discrete gradient operator, see
\cite[p.~258]{Gonzales1992}.

Alternating direction method of multipliers (ADMM) \cite{ADMM2011} is used to solve (\ref{eqn:l2l1anisoll2_eq}). First, we rewrite (\ref{eqn:l2l1anisoll2_eq}) as follows:
\begin{align}\label{formulation_constraint}
\begin{split}
\underset{\mathbf{u}}{\text{min }}&\frac{1}{2}||\mathbf{u}-\mathbf{v}_k||_2^2+
\beta_{1}||\mathbf{s}||_{1}+
\frac{\beta_{2}}{2}||D \mathbf{u}||_2^2+ \iota_\mathbf{w}\\
\text{s.t. } & \mathbf{s} = D \mathbf{u} {\rm \ and \ } \mathbf{w} = \mathbf{u}.
\end{split}
\end{align}
Here $D$ denote the discrete operator of $\nabla$, $D=\begin{pmatrix}[0.6]D_x  \\D_y  \\ \end{pmatrix} \in \mathbb{R}^{2n\times n}$, where $D_x$ and $D_y$ are the first-order difference matrices in the horizontal and vertical directions respectively and $n$ is the number of pixels, $\iota_\mathbf{w}$ is the indicator function, where $\iota_\mathbf{w}=0$ if $\mathbf{w}|_{\Upsilon}=\mathbf{v}_k|_{\Upsilon}$ and $\iota_\mathbf{w}=\infty$ otherwise. Its augmented Lagrangian is given by:
\begin{equation}\label{Largrangian}
L(\mathbf{u},\mathbf{s},\mathbf{w},\boldsymbol{\lambda})=\frac{1}{2}||\mathbf{u}-\mathbf{v}_k||_2^2+\beta_1||\mathbf{s}||_1+\frac{\beta_{2}}{2}||D\mathbf{u}||_2^2+ \iota_\mathbf{w}+\frac{\mu}{2}||E\mathbf{u}-{\bf g}-\boldsymbol{\lambda}||_2^2,
\end{equation}
where $\mu>0$ is a positive constant, $E =
\begin{pmatrix}[0.6]D  \\I  \\ \end{pmatrix}$, ${\bf g}=\begin{pmatrix}[0.6]\mathbf{s}  \\ \mathbf{w}  \\ \end{pmatrix}$ and $\blambda =
\begin{pmatrix}[0.6]\blambda_1  \\\blambda_2  \\ \end{pmatrix}$ the Lagrange multipliers.

The formulation (\ref{Largrangian}) allows us to solve $\mathbf{u}$ and ${\bf g}$
alternately as follows:
\begin{subequations}
\begin{align}
\mathbf{u}^{(r+1)}&=\underset{\mathbf{u}}{\text{argmin }}\ \bigg\{ \frac{1}{2}||\mathbf{u}-\mathbf{v}_k||_2^2+\frac{\beta_{2}}{2}||D\mathbf{u}||_2^2
+\frac{\mu}{2}||E\mathbf{u}-{\bf g}^{(r)}-\blambda^{(r)}||_2^2 \bigg\} \label{sub_u}\\
{\bf g}^{(r+1)} &= \underset{{\bf g}}{\text{argmin }}\ \bigg\{ \beta_1||\mathbf{s}||_1 + \iota_\mathbf{w}+\frac{\mu}{2}||E\mathbf{u}^{(r+1)}-\G-\blambda^{(r)}||_2^2 \bigg\} \label{sub_F} \\
\blambda^{(r+1)} &= \blambda^{(r)}-E\mathbf{u}^{(r+1)}+\G^{(r+1)} \label{sub_sigma}
\end{align}
\end{subequations}
The $\mathbf{u}$-subproblem (\ref{sub_u}) is a least squares problem. Its solution is
\begin{equation}\label{sol_u}
\mathbf{u}^{(r+1)}=(I+\beta_2D^{T}D+\mu E^{T}E)^{-1}(\mathbf{v}_k+\mu E^{T}(\G^{(r)}+\blambda^{(r)})).
\end{equation}
Since periodic boundary conditions are used, the solution can be computed efficiently using the two-dimensional fast Fourier transform (FFT) \cite{Chan2007book}
in $O(n\log n)$ complexity.

For the $\G$-subproblem, the optimal $\mathbf{s}$ and $\mathbf{w}$ can be computed separately as follows:
\begin{equation}\label{sub_s}
\mathbf{s}^{(r+1)} = \underset{\mathbf{s}}{\text{argmin }}\ \bigg\{ \beta_1||\mathbf{s}||_1+\frac{\mu}{2} ||D \mathbf{u}^{(r+1)}-\mathbf{s} - \blambda_1^{(r)}  ||_2^2 \bigg\}
\end{equation}
and
\begin{equation}\label{sub_w}
\mathbf{w}^{(r+1)} = \underset{\mathbf{w}}{\text{argmin }}\ \bigg\{ \iota_\mathbf{w}+\frac{\mu}{2} ||\mathbf{u}^{(r+1)}- \mathbf{w} -\blambda_2^{(r)}||_2^2 \bigg\}
\end{equation}
The solution of (\ref{sub_s}) can be obtained by soft thresholding \cite{Combettes2005}:
\begin{equation}\label{sol_s}
{[\mathbf{s}^{(r+1)}]}_{i}= \text{sgn}([{\bf r}]_i)\cdot \text{max}\{|[{\bf r}]_i|-\frac{\beta_1}{\mu},0\},\;i=1,...,2n,
\end{equation}
where ${\bf r}=D\mathbf{u}^{(r+1)}-\blambda_1^{(r)}$.
The solution of (\ref{sub_w}) is simply
\begin{equation}\label{sol_w}
[\mathbf{w}^{(r+1)}]_i = \left\{
\begin{array}{ll}
[\mathbf{v}_k]_i & {\rm if } \ i \in {\Upsilon} ,\\
{[\mathbf{u}^{(r+1)}-\blambda_2^{(r)}]}_i & {\rm otherwise.}
\end{array}
\right.
\end{equation}
Note that the computation of (\ref{sub_sigma}), (\ref{sol_s}) and (\ref{sol_w}) have a computational complexity of $O(n)$. Hence the computational complexity is $O(n \log n)$ for each iteration.

Our algorithm is summarized in Algorithm \ref{alg:ADMM}. Its convergence to the global minimum is guaranteed by \cite{ADMM2011}.
Once it finishes, we obtain the restored votes $\bf{u}$ for class $k$.
We denote it as $\mathcal{U}_{:,:,k}$.
After the votes for each class are restored
we get a 3D tensor $\mathcal{U}$. The final classification of the $(i,j)$-th pixel is given
by finding the maximum value in $\mathcal{U}_{i,j,:}$, i.e.
$\underset{k}{\text{argmax }}\, \mathcal{U}_{i,j,k}$.

\begin{algorithm}
\begin{centering}
\begin{algorithmic}[1]
\Initialize{Set $r=0$. Choose $\mu>0$, $\mathbf{u}^{(0)}$, $\mathbf{s}^{(0)}$, $\blambda^{(0)}$ and $\mathbf{w}^{(0)}$ where $\mathbf{w}^{(0)}|_{\Upsilon}=\mathbf{v}_k|_{\Upsilon}$.}
\State{\textbf{When stopping criterion is not yet satisfied, do:}}
\State{$\;\;\;\;$ $\mathbf{u}^{(r+1)}\leftarrow(I+\beta_2D^{T}D+\mu E^{T}E)^{-1}(\mathbf{v}_k+\mu E^{T}(\G^{(r)}+\blambda^{(r)}))$}
\State{$\;\;\;\;$ $\mathbf{s}^{(r+1)}\leftarrow \text{sgn}({\bf r})\cdot \text{max}\{|{\bf r}|-\frac{\beta_1}{\mu},0\},$ where$\;{\bf r}=D\mathbf{u}^{(r+1)}-\blambda_1^{(r)}$}
\State{$\;\;\;\;$ $\mathbf{w}^{(r+1)}|_{\Omega\setminus \Upsilon}\leftarrow (\mathbf{u}^{(r+1)}-\blambda_2^{(r)})|_{\Omega\setminus \Upsilon}$}
\State{$\;\;\;\;$ $\blambda^{(r+1)} \leftarrow \blambda^{(r)}-E\mathbf{u}^{(r+1)}+\G^{(r+1)}$}
\end{algorithmic}
\par\end{centering}
\caption{ADMM update process for solving (\ref{eqn:l2l1anisoll2_eq})\label{alg:ADMM}}
\end{algorithm}

We remark that in Stage 1, the operation is along the spectral dimension, i.e. the third index of the tensor,
while in Stage 2, the operation is along the spatial dimension, i.e. the first two indices of the tensor.

\section{Experimental Results}\label{sec:results}

\subsection{Experimental Setup}
\subsubsection{Data Sets}
Three commonly-tested hyperspectral dataset are used in our experiments.
These data sets have pixels labeled so that we can compare the methods
quantitatively.
The first one is the ``Indian Pines" data set acquired by the Airborne Visible/Infrared Imaging Spectrometer (AVIRIS) sensor over the Indian Pines test site in North-western Indiana. It has a spatial resolution of 20 m per pixel and a spectral coverage ranging from 0.2 to 2.4 $\mu$m in 220 spectral bands. However, due to water absorption, 20 of the spectral bands (the 104-108th, 150-163th and 220th bands) are discarded in experiments
in previous papers. Therefore our data set is of size $145 \times 145 \times 200$, and
there are 16 classes in the given ground-truth labels.

The second and third images are the ``University of Pavia" and ``Pavia Center" data sets  acquired by the Reflective Optics System Imaging Spectrometer (ROSIS) sensor over Pavia in northern Italy. The sensor has 1.3 m spatial resolution and spectral coverage ranging from 0.43 to 0.86 $\mu$m. The data set sizes are $610 \times 340 \times 103$ and $1096 \times 715 \times 102$ respectively, where the third dimension is the spectral dimension. Both sets have 9 classes in the ground-truth labels.

\subsubsection{Methods Compared and Parameters Used}

We have compared our method with five well-known classification methods: $\nu$-support vector classifiers ($\nu$-SVC) \cite{Scholkopf2000,Melgani2004} (i.e. the first stage of our method), SVMs with composite kernels (SVM-CK) \cite{Camps2006}, edge-preserving filtering (EPF) \cite{Kang2014}, superpixel-based classification via multiple kernels (SC-MK) \cite{Fang2015G} and multiple-feature-based adaptive sparse representation (MFASR) \cite{Fang2017}. All the tests are run on a laptop computer with an Intel Core i5-7200U CPU, 8 GB RAM and the software platform is MATLAB R2016a.

In the experiments, the parameters are chosen as follows. For the $\nu$-SVC method, the parameters are obtained by performing a five-fold cross-validation \cite{Kohavi1995}. For the SVM-CK method, the parameters are tuned such that it gives the highest classification accuracy.
All parameters of the EPF method, the SC-MK method, and the MFASR method are chosen as stated in \cite{Kang2014,Fang2015G,Fang2017} respectively, except the window size in the EPF method, the number of superpixels and the parameters of the superpixel segmentation algorithm in the SC-MK method, and the sparsity level of the MFASR are tuned such that the highest classification accuracies are obtained.  For our method,  the parameters of the $\nu$-SVC (\ref{nu_svm}) in the first stage are obtained by performing a five-fold cross-validation and the parameters of the optimization problem (\ref{eqn:l2l1anisoll2_eq}) in the second stage are tuned such that it gives the highest classification accuracy.

\subsubsection{Performance Metrics}

To quantitatively evaluate the performance of the methods, we use the
following three widely-used metrics: (i) overall accuracy (OA): the percentage of correctly classified pixels, (ii) average accuracy (AA): the average percentage of correctly classified pixels over each class,  and (iii) kappa coefficient (kappa): the percentage of correctly classified pixels corrected by the number of agreements that would be expected purely by chance \cite{Cohen1960}.

For each method, we perform the classification ten times where each time we randomly choose
a different set of training pixels. In the tables below, we give the averages of these metrics over the ten runs. The accuracies are given in percentage, and the highest accuracy of each category is listed in boldface. In the figures, we count the number of mis-classification for each testing pixel over the ten runs. The numbers of mis-classification are shown in the corresponding heatmap figures, with the heatmap colorbar indicating the number of mis-classifications.

\subsection{Classification Results}\label{sec:comparison_results}

\subsubsection{Indian Pines}\label{sec:comparison_resultsip}
\begin{figure}[h!]
        \centering
            \includegraphics[scale=1.3]{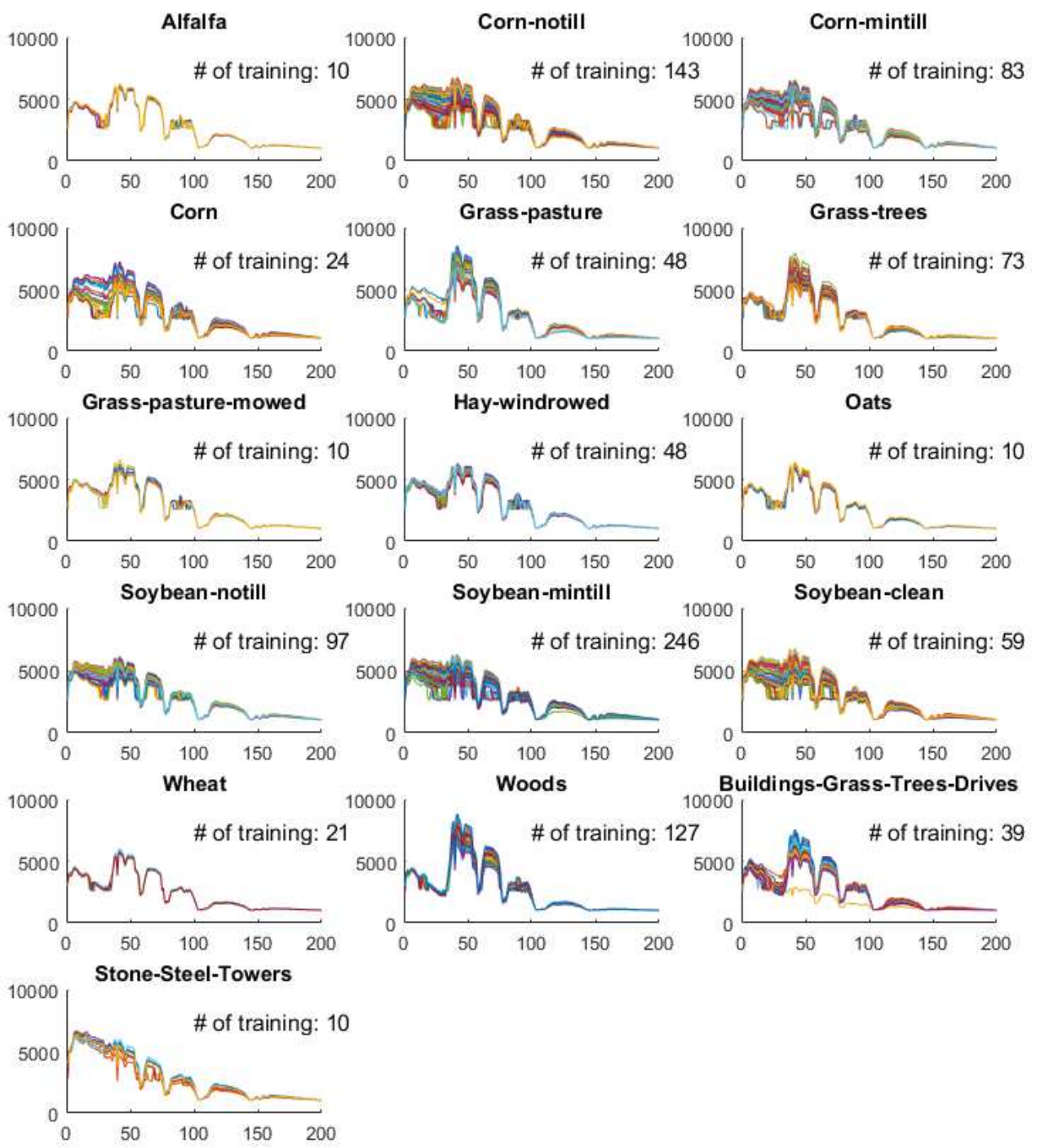}
            \caption[]%
            {{\small Spectra of training pixels of Indian Pines data \label{indianpines_training_sample}}}
        \end{figure}

The Indian Pines data set  consists mainly of big homogeneous regions
and has very similar inter-class spectra (see \Cref{indianpines_training_sample} for the spectra of the training pixels of Indian Pines data where there are three similar classes of corns, three similar classes of grasses and three similar classes of soybeans). It is therefore very difficult to classify it if only spectral information is used. In the experiments, we
choose the same number of training pixels as in \cite{Fang2015,Fang2015G} and
they amount to about 10\% of the pixels from each class.
The rest of the labeled pixels are used as testing pixels.

The number of training and testing pixels as well as the classification accuracies obtained by different methods are reported in \Cref{indianpines_table}. We see that our method generates the best results for all three metrics (OA, AA and kappa) and outperforms the comparing methods by a significant margin. They are at least 0.95\% higher than the others. Also, the second stage of our method improves the overall accuracy of $\nu$-SVC (used in the first stage of our method) by almost 20\%.

\begin{figure}[h!]
 \centering
   \begin{subfigure}[b]{0.25\textwidth}
        \centering
            \includegraphics[width=\textwidth]{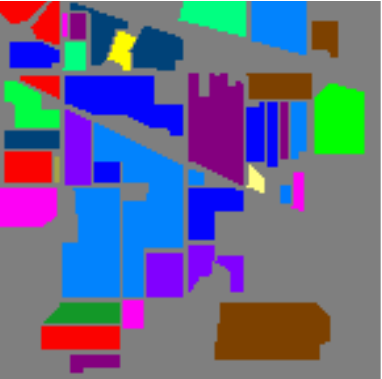}
            \caption[]%
            {{\small Ground Truth}}
        \end{subfigure}
        \hfill
        \begin{subfigure}[b]{0.46\textwidth}
            \centering
            \includegraphics[width=\textwidth]{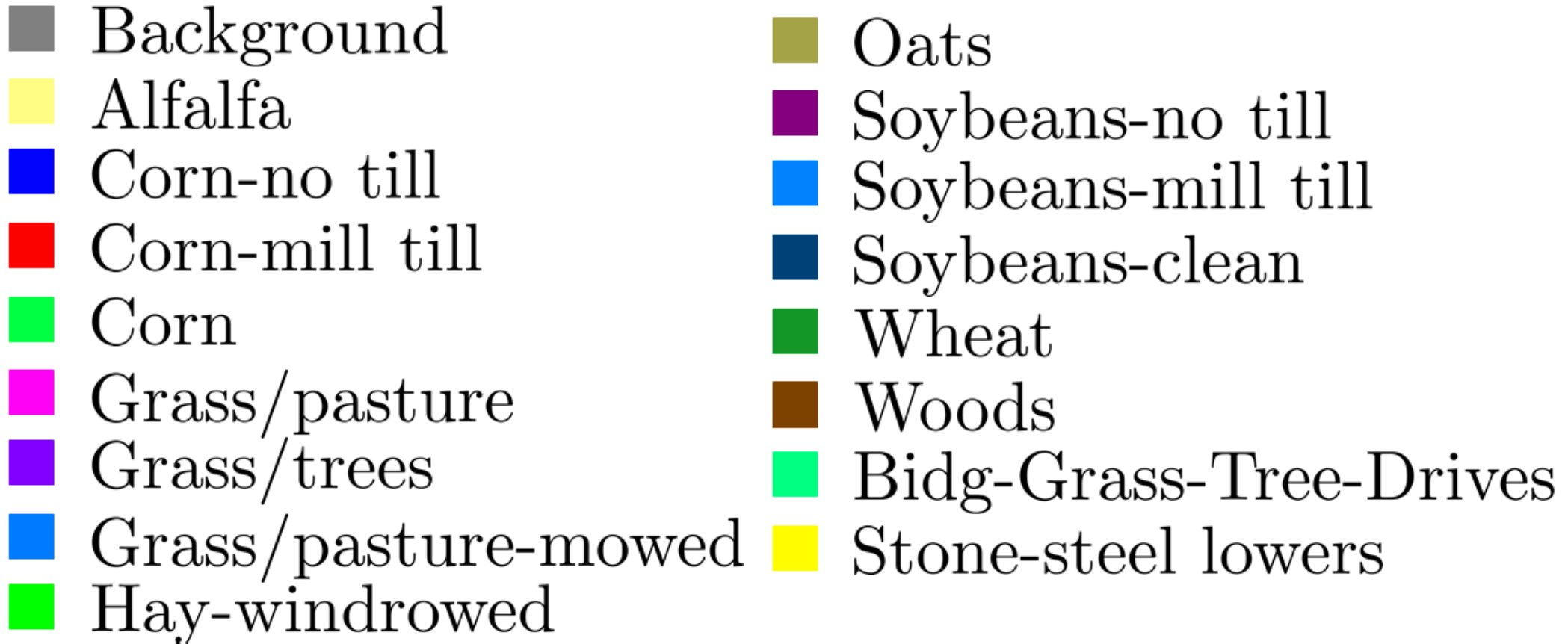}
            \caption[]%
            {{\small Label color}}
        \end{subfigure}
        \hfill
        \begin{subfigure}[b]{0.25\textwidth}
            \centering
            \includegraphics[width=\textwidth]{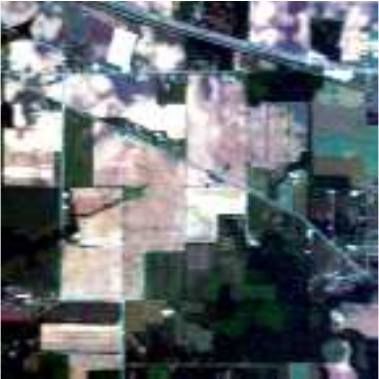}
            \caption[]%
            {{\small False color image}}
        \end{subfigure}
        \hfill
        \vskip\baselineskip
               \begin{subfigure}[b]{0.15\textwidth}
            \centering
            \includegraphics[width=\textwidth]{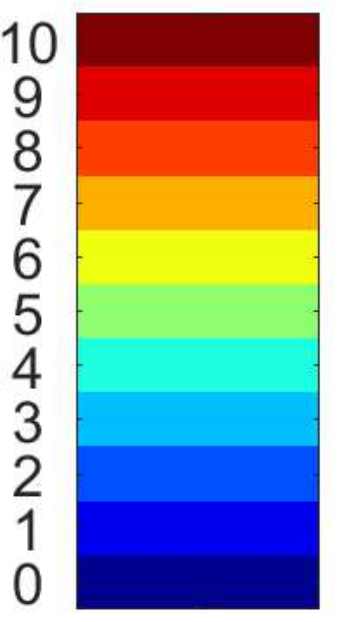}
            \caption[]%
            {{\small Heatmap colorbar}}
        \end{subfigure}
        \hfill
        \centering
        \begin{subfigure}[b]{0.25\textwidth}
            \centering
            \includegraphics[width=\textwidth]{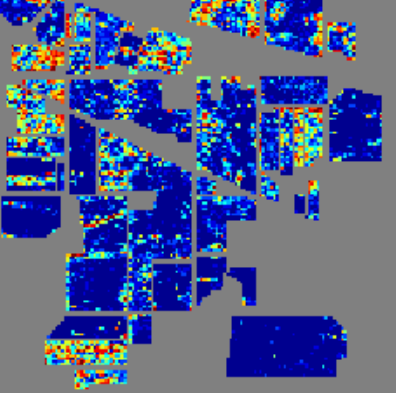}
            \caption[]%
            {{\small SVM \cite{Melgani2004}}}
        \end{subfigure}
        \hfill
        \begin{subfigure}[b]{0.25\textwidth}
            \centering
            \includegraphics[width=\textwidth]{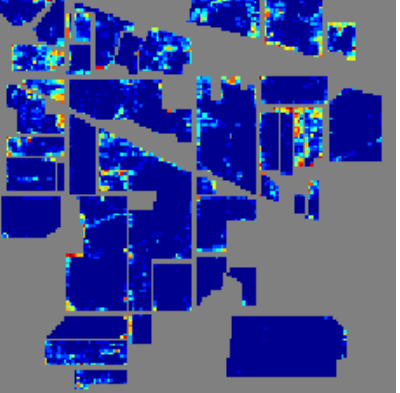}
            \caption[]%
            {{\small SVM-CK \cite{Camps2006}}}
        \end{subfigure}
        \hfill
                \vskip\baselineskip
                \centering
        \begin{subfigure}[b]{0.25\textwidth}
            \centering
            \includegraphics[width=\textwidth]{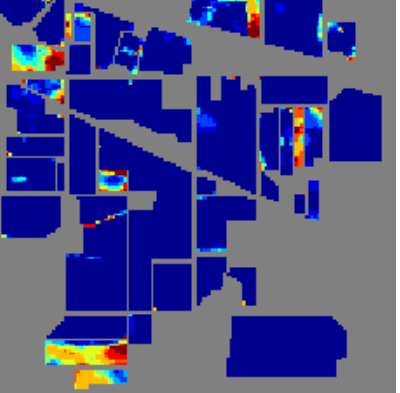}
            \caption[]%
            {{\small EPF \cite{Kang2014}}}
        \end{subfigure}
        \hfill
        \begin{subfigure}[b]{0.25\textwidth}
            \centering
            \includegraphics[width=\textwidth]{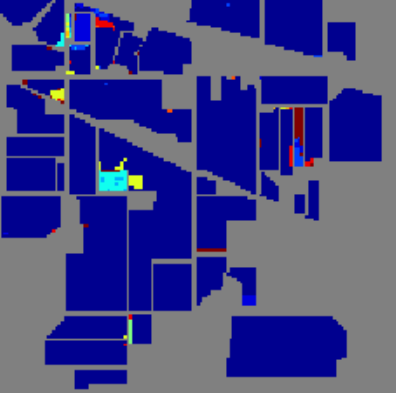}
            \caption[]%
            {{\small SC-MK \cite{Fang2015G}}}
        \end{subfigure}
        \hfill
        \begin{subfigure}[b]{0.25\textwidth}
            \centering
            \includegraphics[width=\textwidth]{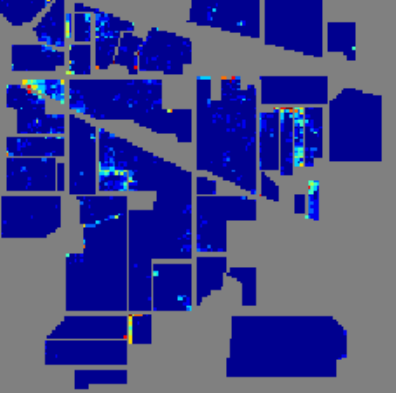}
            \caption[]%
            {{\small MFASR \cite{Fang2017}}}
        \end{subfigure}
        \hfill
                \vskip\baselineskip
                \centering
        \begin{subfigure}[b]{0.25\textwidth}
            \centering
            \includegraphics[width=\textwidth]{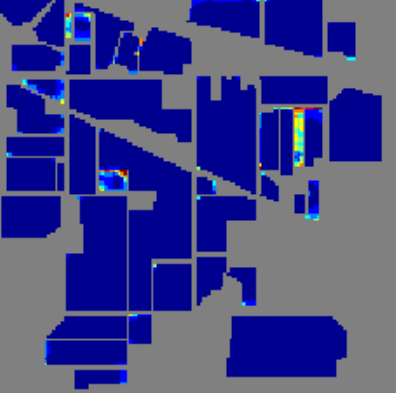}
            \caption[]%
            {{\small Our 2-stage}}
        \end{subfigure}
        \caption{Indian Pines data set. (a) ground-truth labels, (b) label color of the ground-truth labels, (c) false color image, (d) heatmap colorbar, (e)--(j) classification results by different methods. \label{indianpines_fig}}
    \end{figure}

\Cref{indianpines_fig} shows the heatmaps of mis-classifications. The results of the $\nu$-SVC, SVM-CK and EPF methods produce large area of  mis-classifications. The SC-MK also produces mis-classification at the top-right region and the middle-right region which are soybeans-clean and soybeans-no till
respectively. This shows that SC-MK cannot distinguishing these two similar classes well. The heatmap of MFASR method contains scattered regions of mis-classification.  In contrast, our method generates smaller regions of mis-classifications and less errors as it effectively utilizes the spatial information to give an accurate result.

\begin{table}[h!]
\centering
\caption{Number of training/testing pixels and classification accuracies for Indian Pines data set.\label{indianpines_table}}\begin{centering}
%\rotatebox{90}{
\begin{tabular}{|c c|c|c|c|c|c|c|}
\hline
Class &train/test& $\nu$-SVC & SVM-CK & EPF &SC-MK & MFASR & 2-stage \tabularnewline
\hline
Alfalfa&10/36 &  70.28\% & 81.94\%  & 97.29\%& \textbf{100\%}& 98.06\% & 99.17\%\tabularnewline
\hline
Corn-no till&143/1285 &  77.90\% & 89.98\% & 96.03\% & 95.44\%& 96.66\% &\textbf{97.89\%} \tabularnewline
\hline
Corn-mill till&83/747 &  67.80\% & 89.68\% & 97.75\% & 97.16\%& 97.94\% &\textbf{98.73\%}\tabularnewline
\hline
Corn & 24/213& 52.96\% & 86.24\% &93.03\% &\textbf{99.25\%}&91.69\% &99.01\%\tabularnewline
\hline
Grass/pasture &48/435&  89.13\% & 93.31\% & \textbf{99.17\%}& 96.67\%& 94.62\%&96.92\% \tabularnewline
\hline
Grass/trees &73/657& 96.15\% & 98.98\% & 96.02\%& 99.70\%& 99.56\% &\textbf{99.74\%}\tabularnewline
\hline
Grass/pasture-mowed&10/18 & 93.33\% & 96.11\% & 99.47\%& \textbf{100\%}& \textbf{100\%}&\textbf{100\%}  \tabularnewline
\hline
Hay-windrowed&48/430 & 93.93\% &98.42\% & \textbf{100\%}& \textbf{100\%}& 99.98\% &\textbf{100\%}\tabularnewline
\hline
Oats&10/10 & 90.00\% & \textbf{100\%} & 96.25\%& \textbf{100\%}& \textbf{100\%}&\textbf{100\%} \tabularnewline
\hline
Soybeans-no till&97/875 & 72.26\% & 88.81\% & 92.21\%& 94.62\%& \textbf{96.03\%}&96.01\% \tabularnewline
\hline
Soybeans-mill till&246/2209 & 79.71\% & 91.57\% & 86.65\%& \textbf{98.80\%}& 98.58\%&99.54\% \tabularnewline
\hline
Soybeans-clean &59/534& 67.66\% & 85.90\% & 96.26\% & 96.29\%& 97.06\% &\textbf{99.64\%}\tabularnewline
\hline
Wheat &21/184& 96.09\% & 98.64\%& \textbf{100\%} & 99.67\%& 99.57\% &\textbf{100\%}\tabularnewline
\hline
Woods &127/1138& 91.89\% & 96.85\% & 95.24\%& \textbf{99.99\%}& 99.89\%&99.91\% \tabularnewline
\hline
Bridg-Grass-Tree-Drives&39/347 & 56.97\% & 88.01\% & 93.70\% & 98.39\%& 98.01\%&\textbf{99.14\%}  \tabularnewline
\hline
Stone-steel lowers &10/83& 85.66\% & 98.43\% & 96.11\% & 97.71\%& \textbf{98.92\%} &96.39\% \tabularnewline
\hline
\hline
OA & &79.78\% & 92.11\% & 93.34\%& 97.83\%& 97.88\% &\textbf{98.83\%}\tabularnewline
\hline
AA & &80.11\% & 92.68\% &  95.95\% & 98.35\%& 97.91\%&\textbf{98.88\%}\tabularnewline
\hline
kappa & & 0.769 & 0.910 &  0.924 & 0.975& 0.976&\textbf{0.987}\tabularnewline
\hline
\end{tabular}

\par\end{centering}
\end{table}

\subsubsection{University of Pavia}\label{sec:comparison_resultsup}
The University of Pavia data set consists of regions with various shapes, including thin and thick structures and large homogeneous regions. Hence it can be used to test the ability of the classification methods on handling different shapes. In the experiments, we choose the same number of training pixels (200 for each class) as in \cite{Fang2015G}. This accounts for approximately 4\% of the labeled pixels. 
The remaining ones are used as testing pixels.

\begin{figure}[h!]
\centering
\begin{subfigure}[b]{0.2\textwidth}
            \centering
            \includegraphics[width=\textwidth]{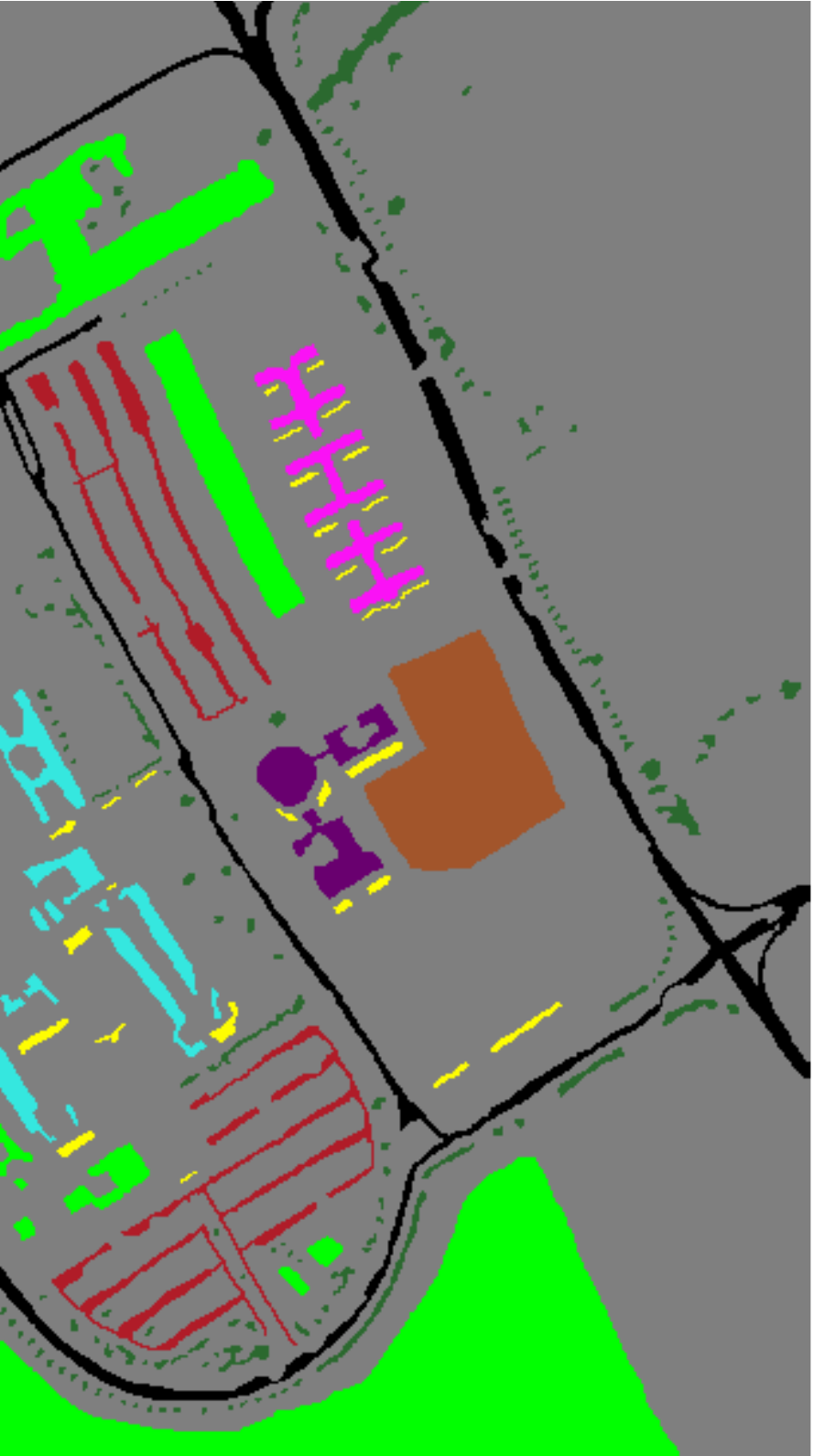}
            \caption[]%
            {{\small Ground Truth}}
        \end{subfigure}
        \hfill
        \begin{subfigure}[b]{0.23\textwidth}
            \centering
            \includegraphics[width=\textwidth]{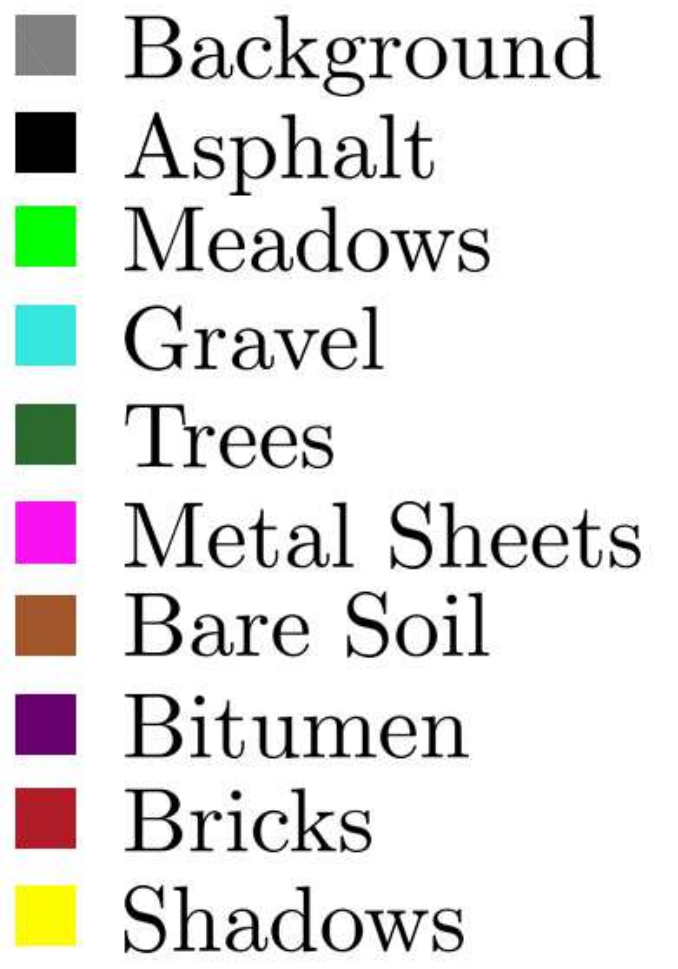}%z_PaviaU_label}
            \caption[]%
            {{\small Label color}}
        \end{subfigure}
        \hfill
        \begin{subfigure}[b]{0.2\textwidth}
            \centering
            \includegraphics[width=\textwidth]{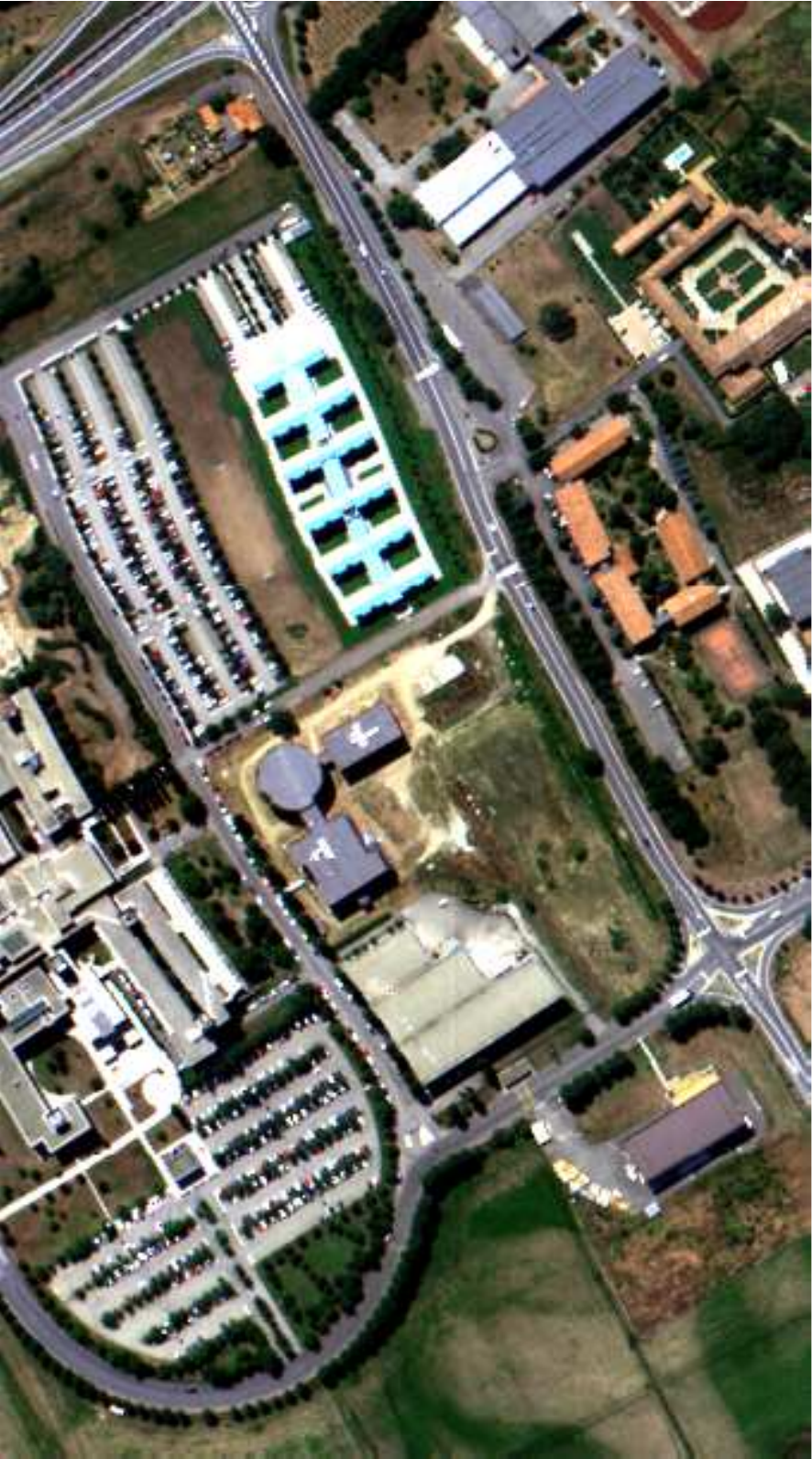}
            \caption[]%
            {{\small False color image}}
        \end{subfigure}
        \hfill
        \vskip\baselineskip
 \centering
 \begin{subfigure}[b]{0.2\textwidth}
            \centering
            \includegraphics[width=\textwidth]{figures/heatmap_bar_thick.pdf}
            \caption[]%
            {{\small Heatmap colorbar}}
        \end{subfigure}
        \hfill
        \begin{subfigure}[b]{0.2\textwidth}
            \centering
            \includegraphics[width=\textwidth]{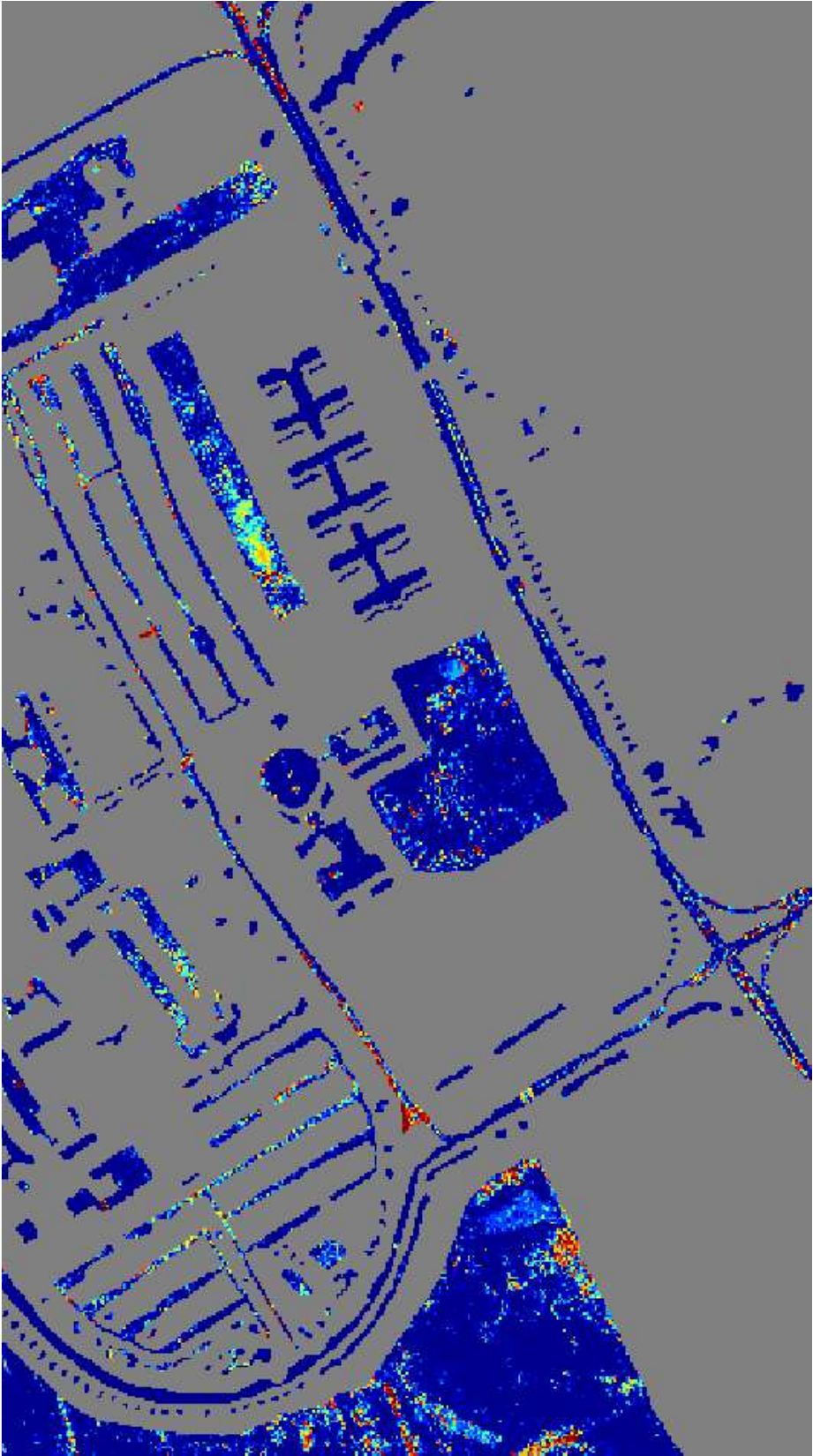}
            \caption[]%
            {{\small $\nu$-SVC \cite{Scholkopf2000,Melgani2004}}}
        \end{subfigure}
        \hfill
        \begin{subfigure}[b]{0.2\textwidth}
            \centering
            \includegraphics[width=\textwidth]{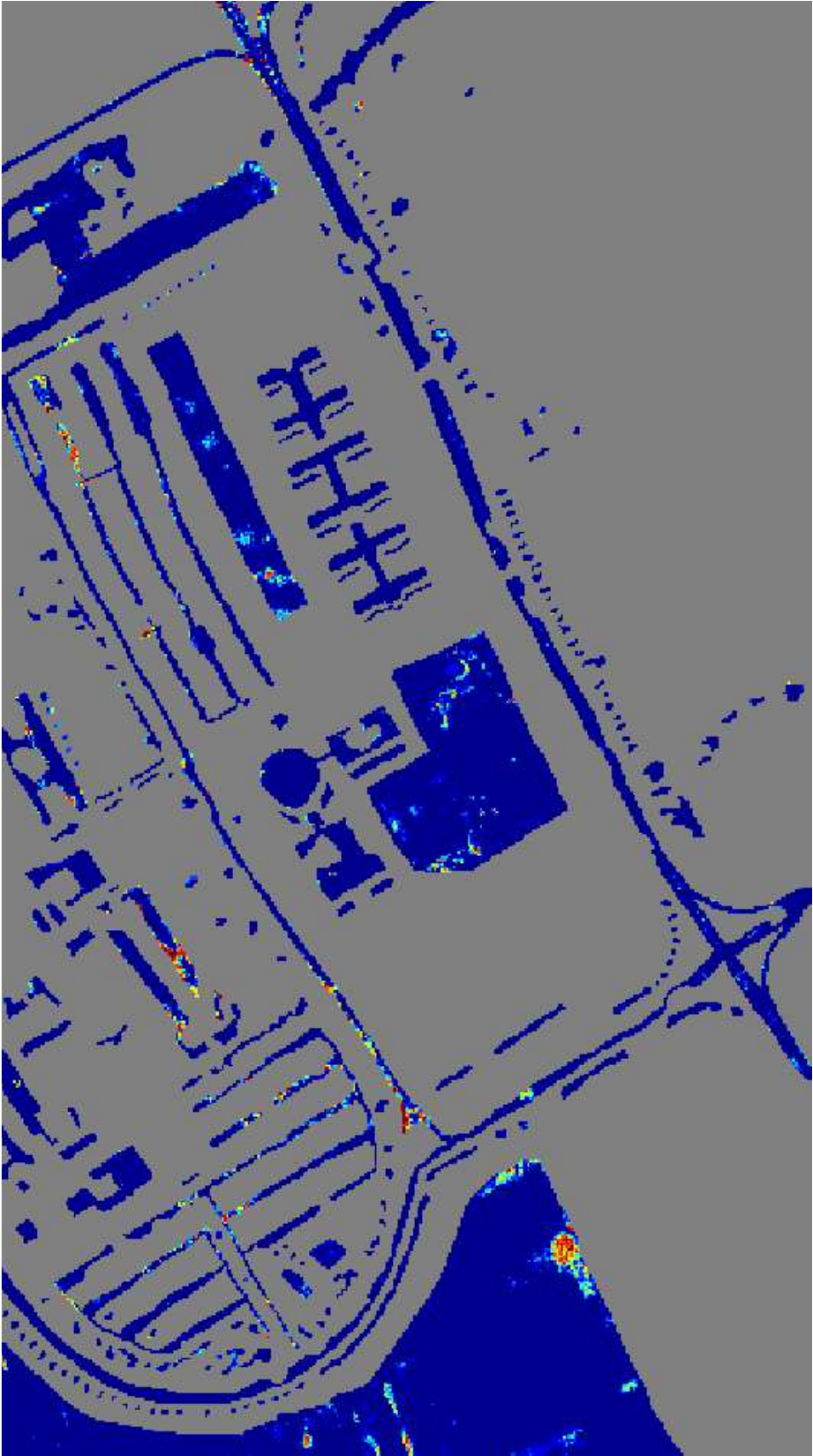}
            \caption[]%
            {{\small SVM-CK \cite{Camps2006}}}
        \end{subfigure}
        \hfill
        \begin{subfigure}[b]{0.2\textwidth}
            \centering
            \includegraphics[width=\textwidth]{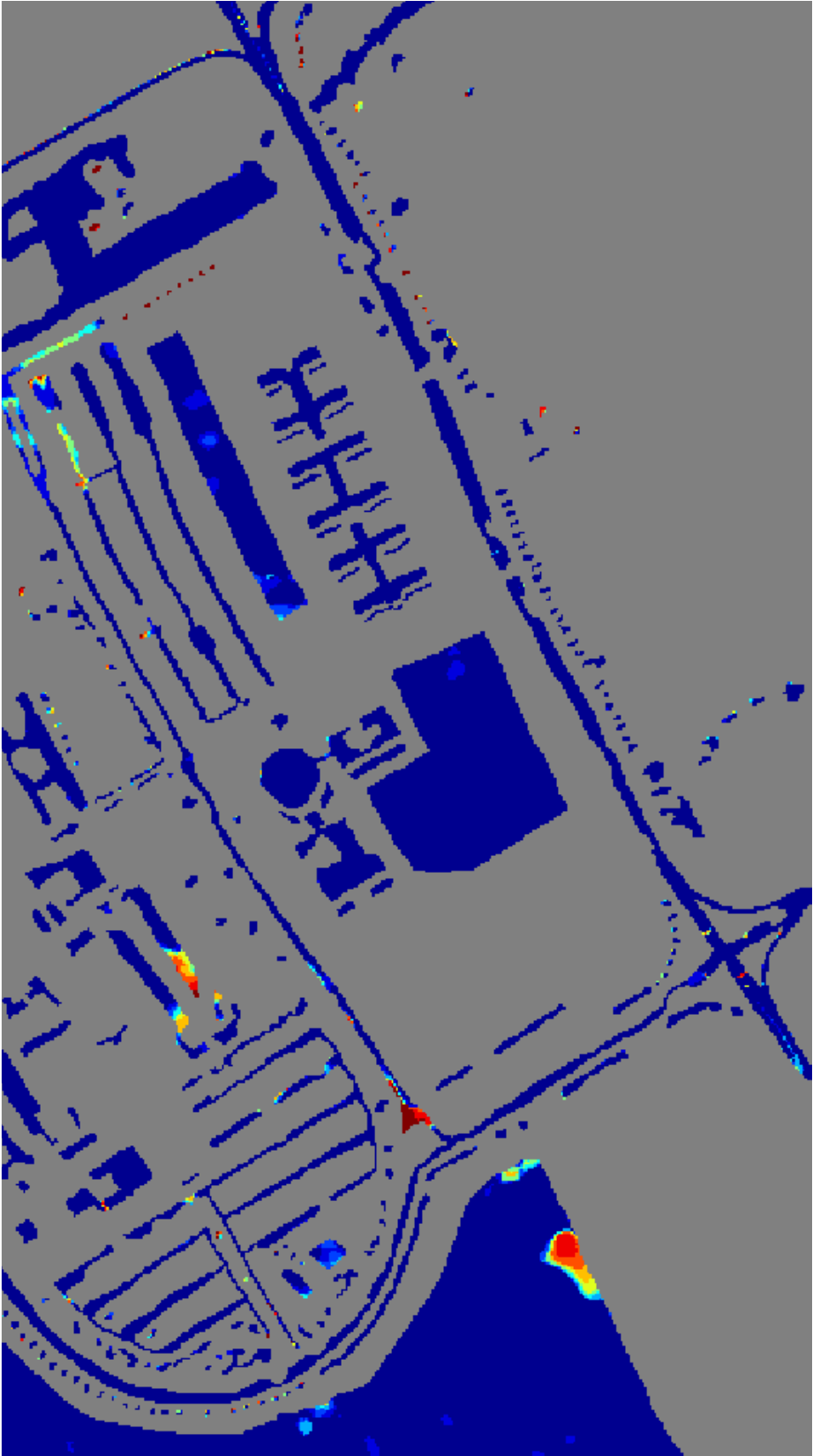}
            \caption[]%
            {{\small EPF \cite{Kang2014}}}
        \end{subfigure}
        \hfill
                        \vskip\baselineskip
        \centering
        \begin{subfigure}[b]{0.2\textwidth}
            \centering
            \includegraphics[width=\textwidth]{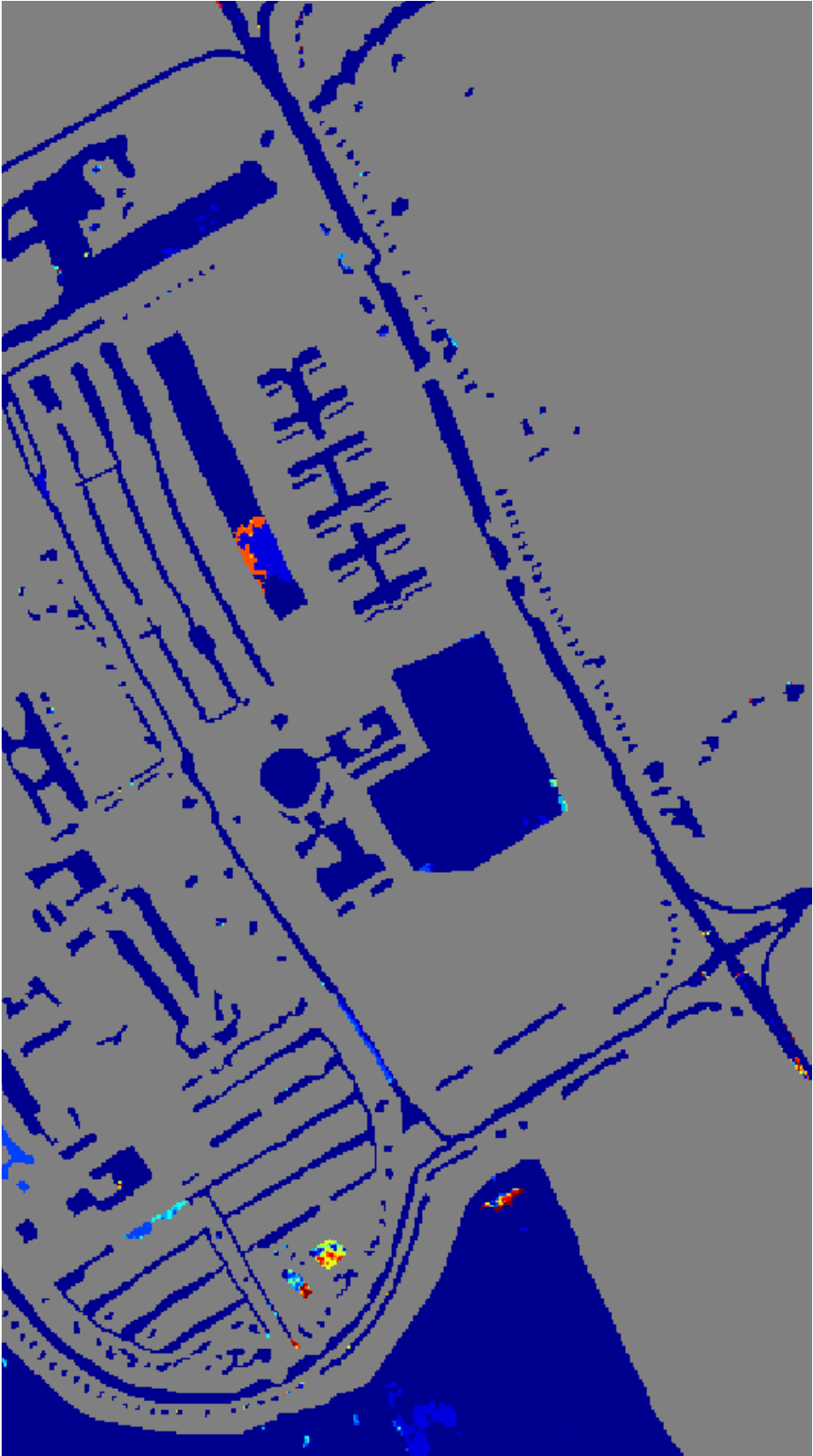}
            \caption[]%
            {{\small SC-MK \cite{Fang2015G}}}
        \end{subfigure}
        \hfill
        \begin{subfigure}[b]{0.2\textwidth}
            \centering
            \includegraphics[width=\textwidth]{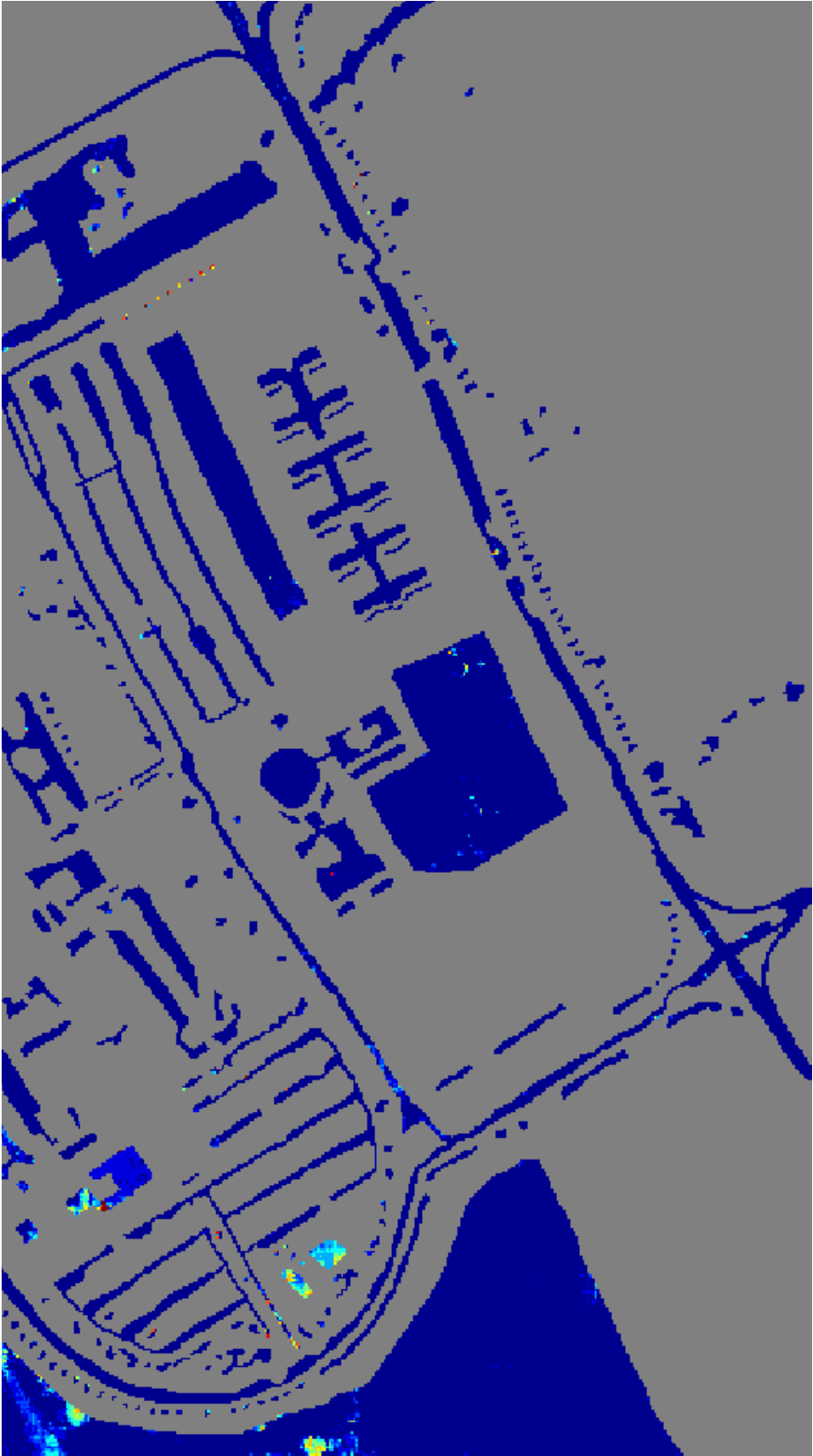}
            \caption[]%
            {{\small MFASR \cite{Fang2017}}}
        \end{subfigure}
        \hfill
        \begin{subfigure}[b]{0.2\textwidth}
            \centering
            \includegraphics[width=\textwidth]{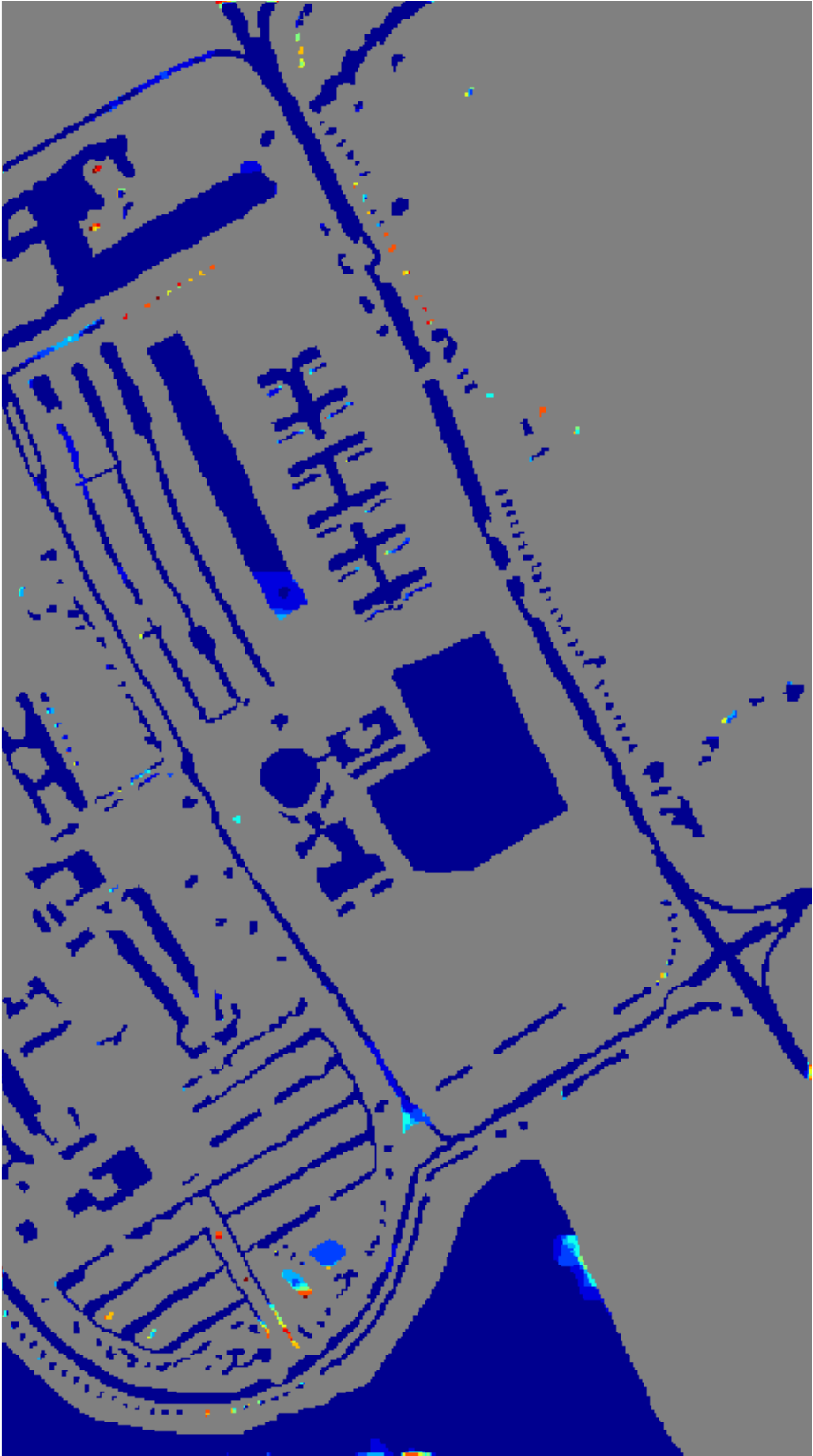}
            \caption[]%
            {{\small Our 2-stage}}
        \end{subfigure}
        \caption{University of Pavia data set. (a) ground-truth labels, (b) label color of the ground-truth labels, (c) false color image, (d) heatmap colorbar, (e)--(j) classification results by different methods. \label{paviaU_fig}}
    \end{figure}

\Cref{paviaU_table} reports the classification accuracies obtained by different methods.
We see that the performances of SC-MK, MFASR, and our method are very close: approximately 99\% in all three metrics (OA, AA and kappa) and they outperform the $\nu$-SVC, SVM-CK and EPF methods. \Cref{paviaU_fig} shows the heatmaps of mis-classifications. The $\nu$-SVC, SVM-CK and EPF methods produce large regions of mis-classifications. The SC-MK method produces many mis-classifications at the middle and bottom regions where the meadows are. The MFASR method and our method generate smaller regions
of mis-classification.

\begin{table}[h!]
\centering
\caption{Number of training/testing pixels and classification accuracies for University of Pavia data set.\label{paviaU_table}}
\begin{tabular}{|c c|c|c|c|c|c|c|}
\hline
Class &train/test& $\nu$-SVC & SVM-CK & EPF &SC-MK & MFASR & 2-stage\tabularnewline
\hline
Asphalt & 200/6431 & 84.65\% & 95.84\% & 98.84\%& 99.06\%& \textbf{99.44\%} & 98.68\% \tabularnewline
\hline
Meadows & 200/18449 & 89.96\% & 97.62\% & 99.62\%& 98.14\%& 98.52\% & \textbf{98.78\%} \tabularnewline
\hline
Gravel  &200/1899& 83.59\%& 91.99\% & 95.50\%& \textbf{99.98\%}& 99.80\%&99.69\% \tabularnewline
\hline
Trees &200/2864 & 94.94\%&97.95\% & 98.94\%& \textbf{99.03\%}& 98.02\% &96.56\%\tabularnewline
\hline
Metal Sheets&200/1145 & 99.59\%& 99.97\%& 99.03\% & 99.87\%& 99.91\%& \textbf{100\%} \tabularnewline
\hline
Bare Soil&200/4829 & 90.69\%& 97.49\% & 92.95\%& 99.70\%& 99.78\% &\textbf{100\%}\tabularnewline
\hline
Bitumen &200/1130& 92.73\%& 98.41\% & 93.84\%& 100\%& 99.92\% &\textbf{100\%}\tabularnewline
\hline
Bricks&200/3482 & 82.59\%& 92.71\% & 92.92\%& 99.05\%& \textbf{99.41\%}&99.02\%\tabularnewline
\hline
Shadows &200/747& 99.60\%& 99.92\% & 99.30\%& 99.99\%& \textbf{100\%}&99.18\% \tabularnewline
\hline
\hline
OA& & 89.16\% & 96.80\% & 97.60\%& 98.83\%& \textbf{99.02\%}&98.89\%\tabularnewline
\hline
AA& & 90.93\% & 96.88\% & 96.77\%& \textbf{99.42\%}& \textbf{99.42\%}&99.10\%
\tabularnewline
\hline
kappa & & 0.857 & 0.957 &  0.968 & 0.984& \textbf{0.987} &0.985\tabularnewline
\hline
\end{tabular}
\end{table}

\subsubsection{Pavia Center}\label{sec:comparison_resultspc}
\begin{figure}[h!]
 \centering
  \begin{subfigure}[b]{0.32\textwidth}
            \centering
            \includegraphics[width=\textwidth]{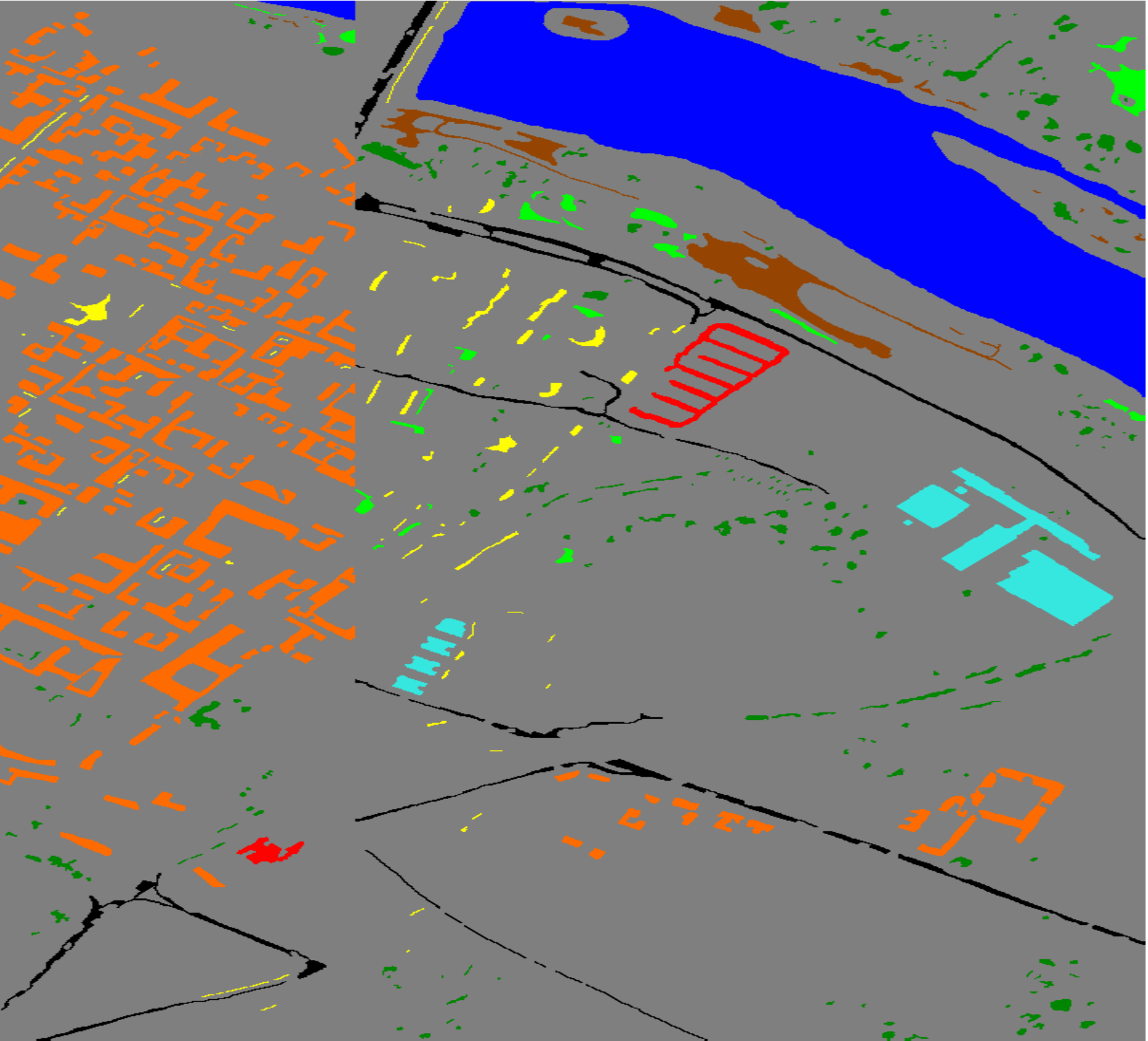}
            \caption[]%
            {{\small Ground Truth}}
        \end{subfigure}
        \hfill
        \begin{subfigure}[b]{0.125\textwidth}
            \centering
            \includegraphics[width=\textwidth]{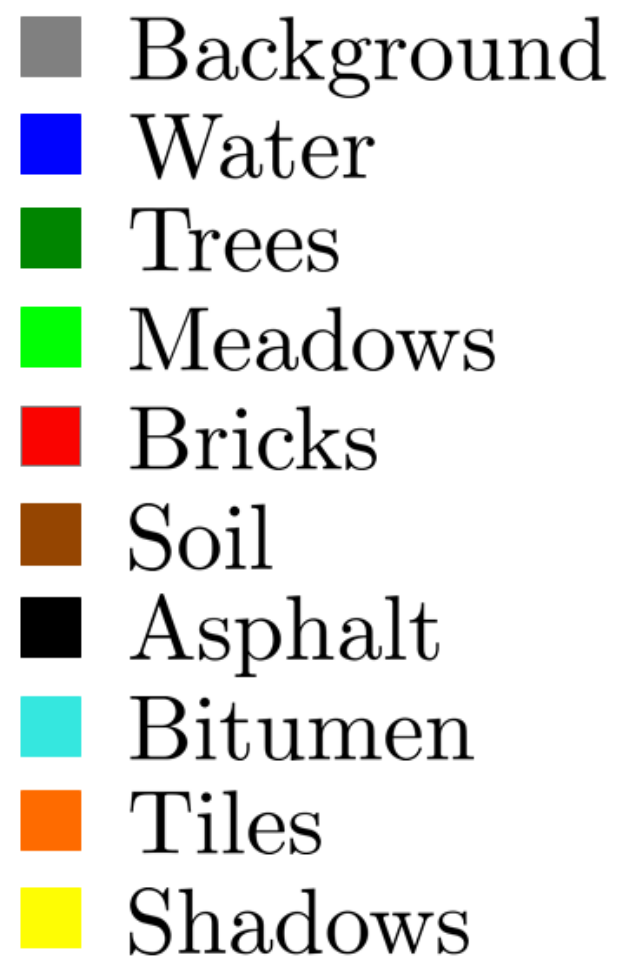}
            \caption[]%
            {{\small Label color}}
        \end{subfigure}
        \hfill
        \begin{subfigure}[b]{0.32\textwidth}
            \centering
            \includegraphics[width=\textwidth]{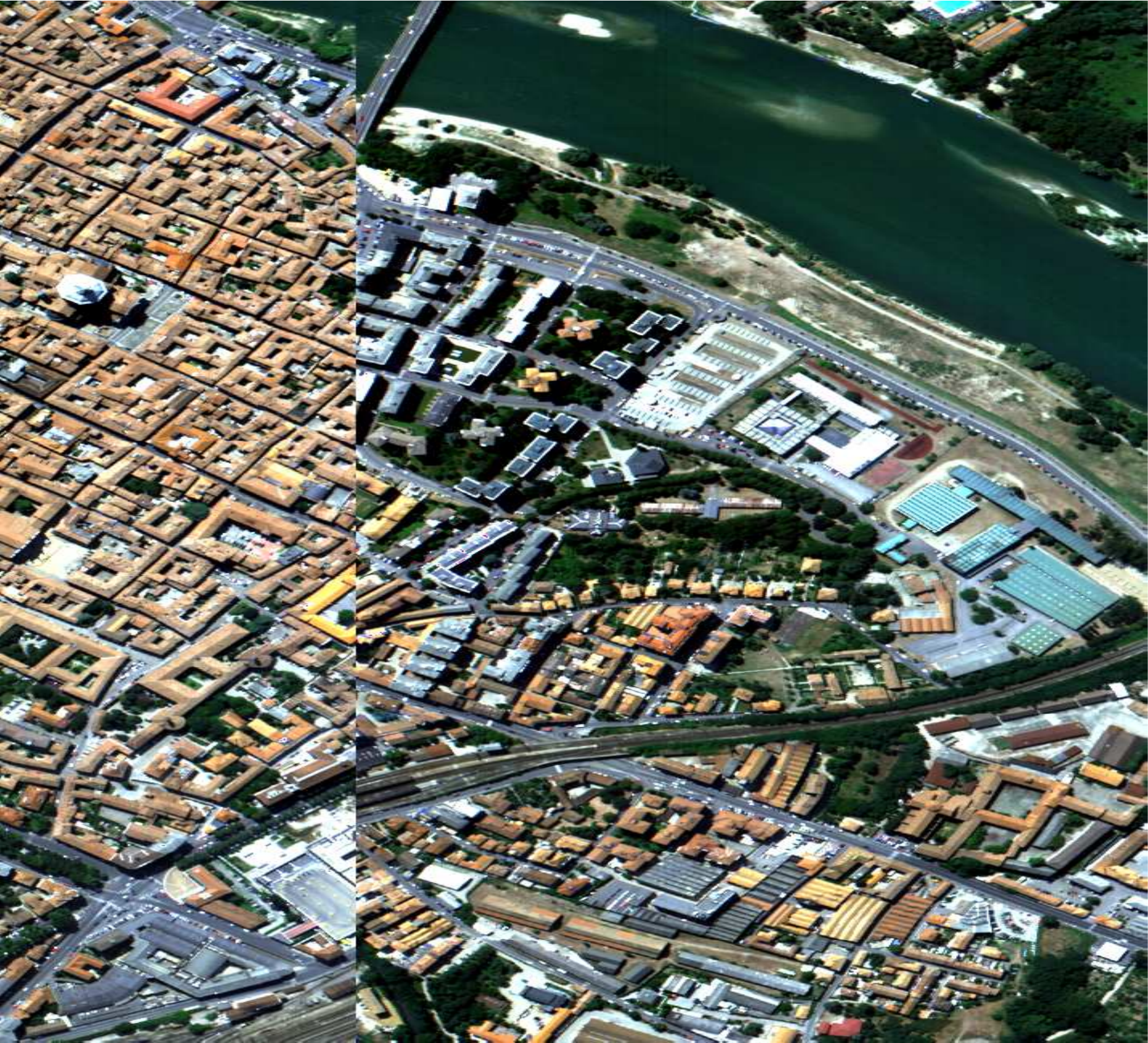}
            \caption[]%
            {{\small False color image}}
        \end{subfigure}
        \hfill
         \begin{subfigure}[b]{0.125\textwidth}
            \centering
            \includegraphics[width=\textwidth]{figures/heatmap_bar_thick.pdf}
            \caption[]%
            {{\small Heatmap colorbar}}
        \end{subfigure}
        \hfill
        \vskip\baselineskip
        \centering
        \begin{subfigure}[b]{0.32\textwidth}
            \centering
            \includegraphics[width=\textwidth]{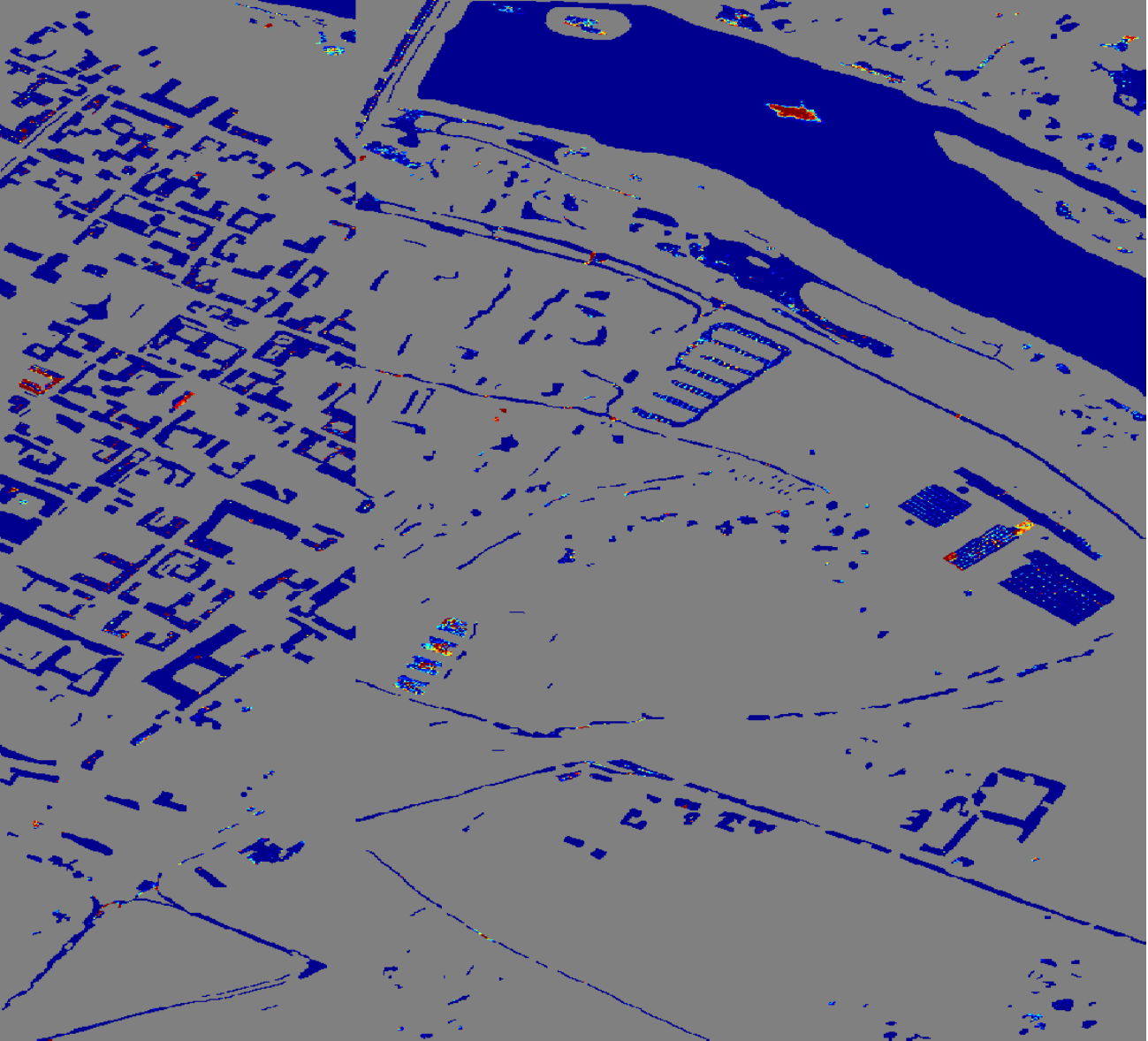}
            \caption[]%
            {{\small $\nu$-SVC \cite{Scholkopf2000,Melgani2004}}}
        \end{subfigure}
        \hfill
        \begin{subfigure}[b]{0.32\textwidth}
            \centering
            \includegraphics[width=\textwidth]{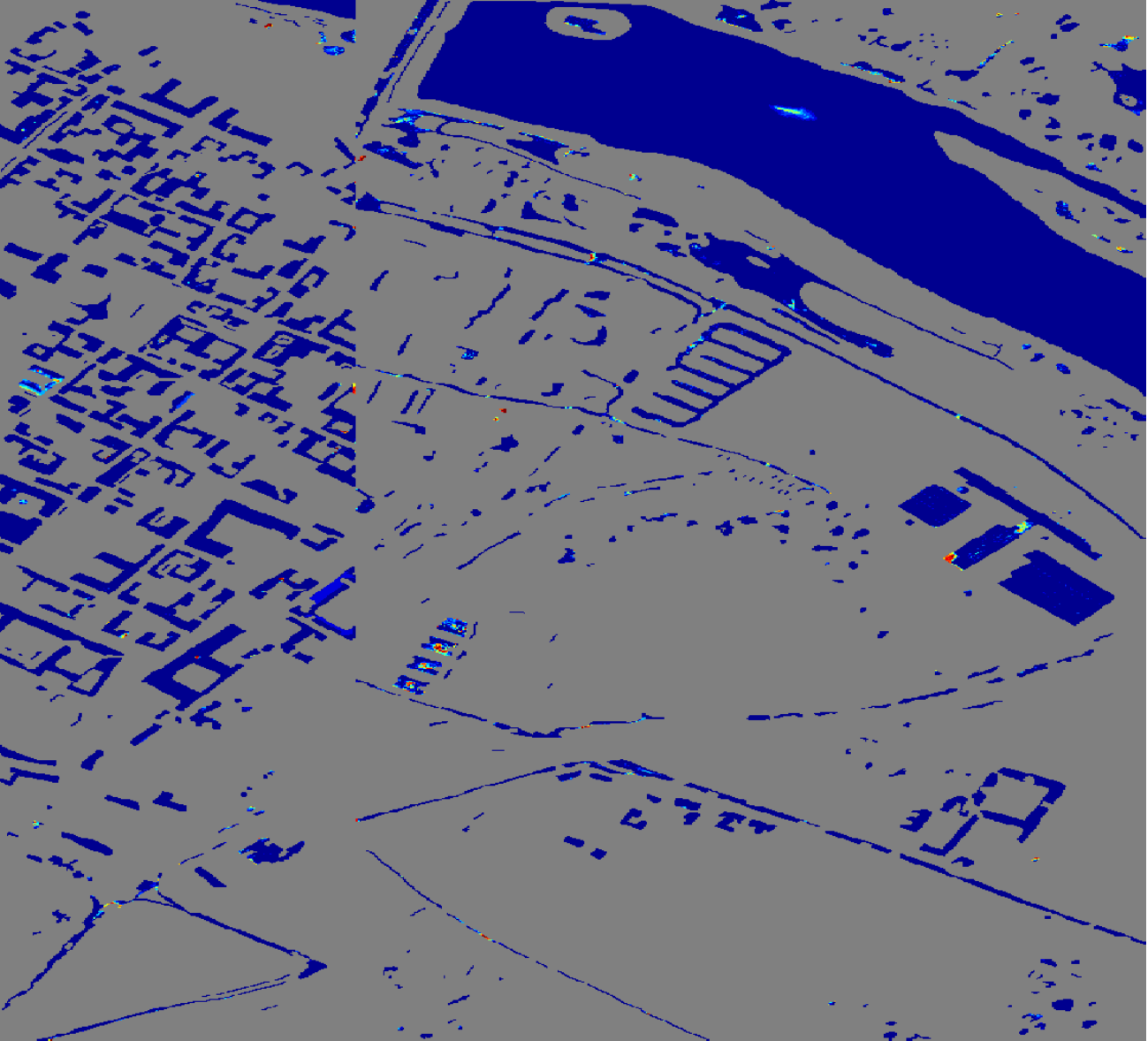}
            \caption[]%
            {{\small SVM-CK \cite{Camps2006}}}
        \end{subfigure}
        \hfill
        \begin{subfigure}[b]{0.32\textwidth}
            \centering
            \includegraphics[width=\textwidth]{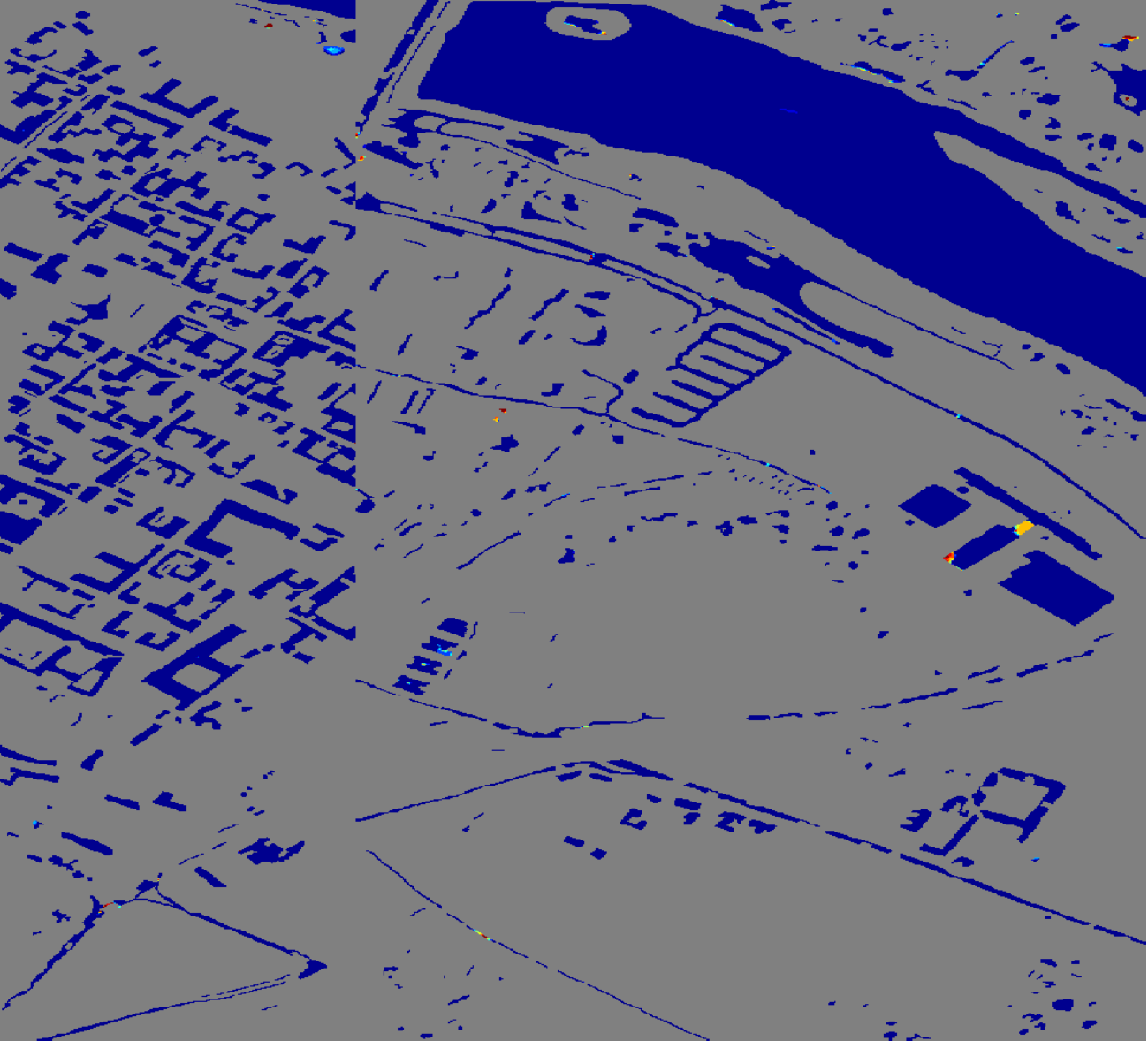}
            \caption[]%
            {{\small EPF  \cite{Kang2014}}}
        \end{subfigure}
        \vskip\baselineskip
        \begin{subfigure}[b]{0.32\textwidth}
            \centering
            \includegraphics[width=\textwidth]{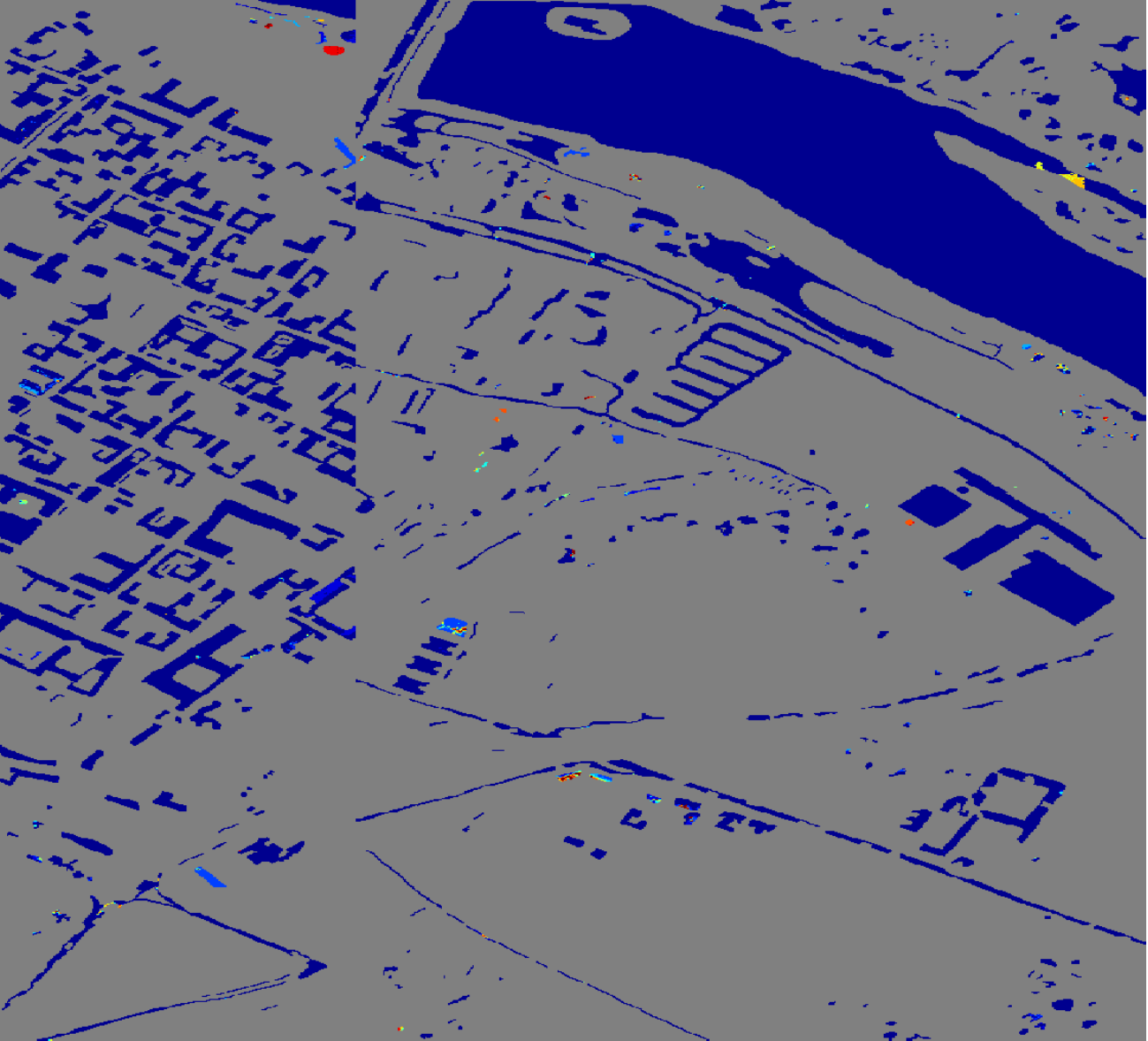}
            \caption[]%
            {{\small SC-MK  \cite{Fang2015G}}}
        \end{subfigure}
        \hfill
        \begin{subfigure}[b]{0.32\textwidth}
            \centering
            \includegraphics[width=\textwidth]{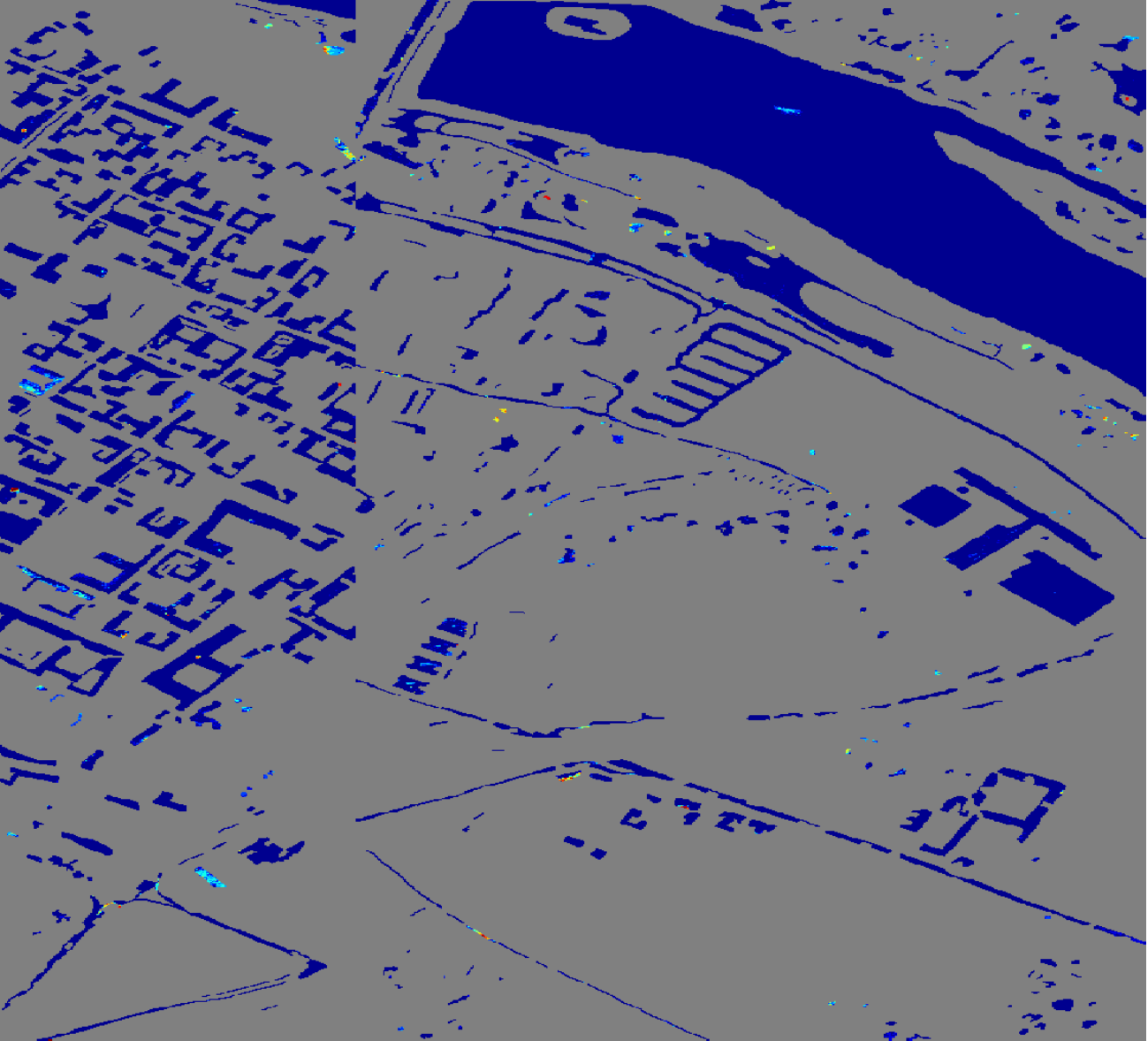}
            \caption[]%
            {{\small MFASR \cite{Fang2017}}}
        \end{subfigure}
        \hfill
        \begin{subfigure}[b]{0.32\textwidth}
            \centering
            \includegraphics[width=\textwidth]{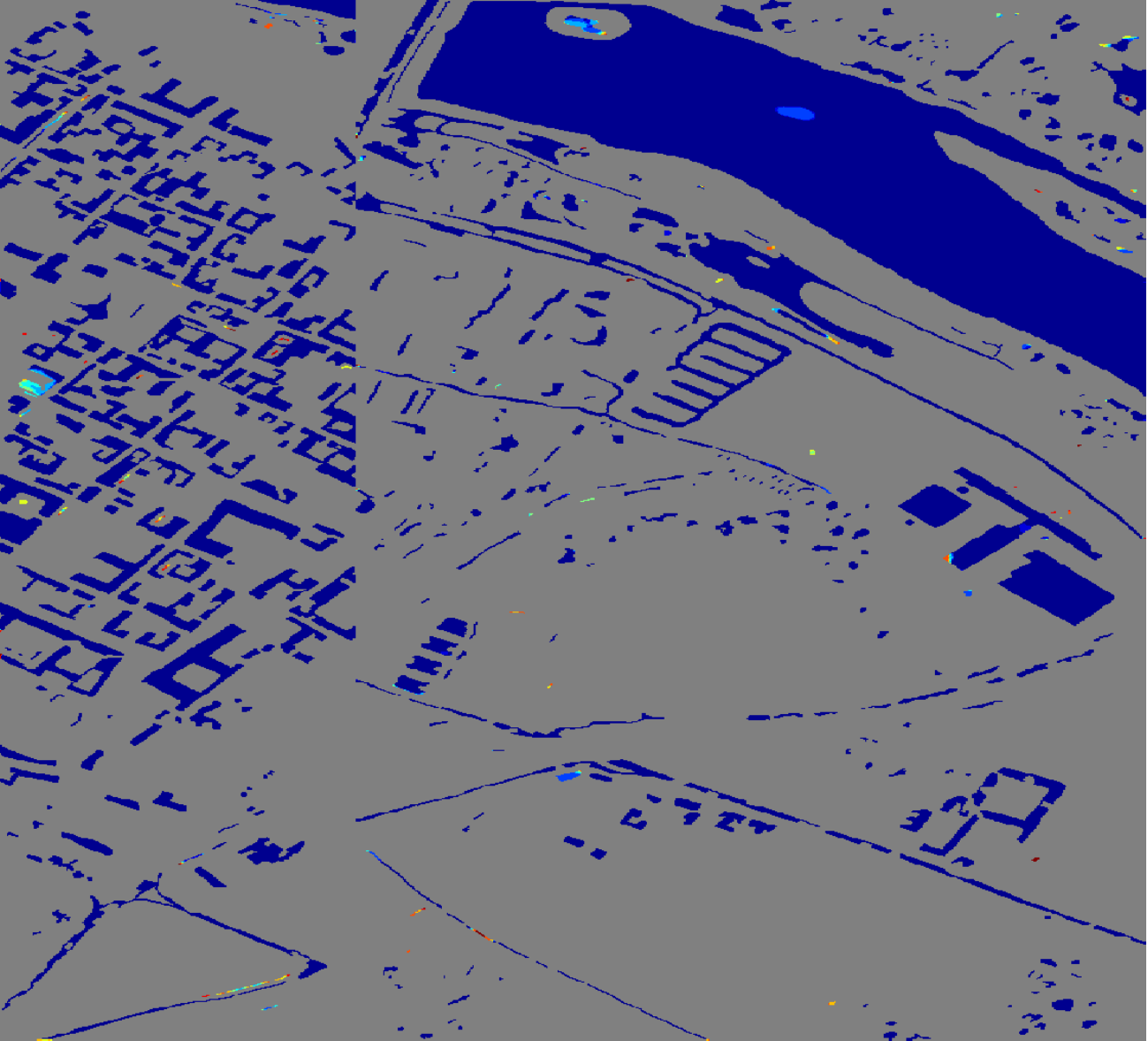}
            \caption[]%
            {{\small Our 2-stage}}
        \end{subfigure}
        \caption{Pavia Center data set. (a) ground-truth labels, (b) label color of the ground-truth labels, (c) false color image, (d) heatmap colorbar, (e)--(j) classification results by different methods. \label{paviacenter_fig}}
    \end{figure}

The Pavia Center data set also consists of regions with various shapes. In the experiments, we use the same number of training pixels as in \cite{Liu2017}
(150 training pixels per class). This accounts for approximately 1\% of the labeled pixels. The rest of the labeled pixels are used as testing pixels. \Cref{paviacenter_table} reports the number of training/testing pixels and the classification accuracies of different methods.
We see that the EPF method gives the highest OA and kappa while our method gives the second highest and their values differ by about 0.1\%. However, our method gives the highest AA (99.12\%) which outperforms the EPF method by almost 1\%. The SC-MK and MFASR methods give slightly worse accuracies than our method.
\Cref{paviacenter_fig} shows the heatmaps of mis-classifications.

\begin{table}[h!]
\centering
%\begin{center}
\caption{Number of training/testing pixels and classification accuracies for Pavia Center data set.\label{paviacenter_table}}
\begin{tabular}{|c c|c|c|c|c|c|c|}
\hline
Class &train/test& $\nu$-SVC & SVM-CK & EPF &SC-MK & MFASR & 2-stage\tabularnewline
\hline
Water & 150/65128 & 99.54\% & 99.82\% & \textbf{100\%}& 99.86\%& 99.97\% & 99.66\% \tabularnewline
\hline
Trees & 150/6357 & 94.22\% & 95.61\% & \textbf{99.11\%}& 94.59\%& 95.52\% & 98.61\%\tabularnewline
\hline
Meadows  &150/2741& 95.14\%& 96.15\% & 97.16\%& 98.78\%& 98.54\% & \textbf{98.84\%}\tabularnewline
\hline
Bricks &150/2002 & 92.56\%&97.37\% & 90.08\%& 99.91\%& 99.62\% & \textbf{99.98\%}\tabularnewline
\hline
Soil&150/6399 & 94.31\%& 96.51\%& 99.40\% & \textbf{99.76\%}& 99.59\%& 98.69\%\tabularnewline
\hline
Asphalt&150/7375 & 95.94\%& 97.34\% & 98.86\%& 99.24\%& 98.76\% & \textbf{99.60\%}\tabularnewline
\hline
Bitumen &150/7137& 89.99\%& 94.75\% & \textbf{99.79\%}& 98.64\%& 99.55\% & 97.86\%\tabularnewline
\hline
Tiles&150/2972 &97.42\%& 99.33\% & \textbf{99.97\%}& 99.32\%& 99.05\%&99.52\%\tabularnewline
\hline
Shadows &150/2015& 99.98\%& \textbf{100\%} & 99.96\%& 99.85\%& 99.97\% & 99.27\%\tabularnewline
\hline
\hline
OA& & 97.54\% & 98.80\% & \textbf{99.59\%}& 99.31\%& 99.33\% & 99.42\%\tabularnewline
\hline
AA& & 95.46\% & 97.43\% & 98.26\%& 98.88\%& 98.95\%&\textbf{99.12\%}
\tabularnewline
\hline
kappa & & 0.965 & 0.983 &  \textbf{0.994} & 0.990& 0.990&0.991\tabularnewline
\hline
\end{tabular}
\end{table}

\subsection{Advantages of Our 2-stage Method}

\subsubsection{Percentage of Training Pixels}

Since our method improves on the classification accuracy by using the
spatial information, it is expected to be a better method if the training
percentage (percentage of training pixels) is higher. To verify that,
\Cref{trainper_indianpines,trainper_paviaU,trainper_paviacenter} show the overall accuracies
obtained by our method on the three data sets with different levels of training percentage.
We see that our method outperforms the other methods when training percentage is high. When it is not high, our method still gives a classification accuracy that is close to the best method compared.

\begin{table}[h!]
\centering
%\begin{center}
\caption{Classification results on the Indian Pines data with different levels of training pixels.\label{trainper_indianpines}}
\begin{tabular}{|c|c|c|c|c|}
\hline
Method \textbackslash Training percentage & 5\% & 10\% & 20\% &40\% \tabularnewline
\hline
$\nu$-SVC  & 73.49\% & 79.78\% & 84.98\% & 88.55\%
\tabularnewline
\hline
SVM-CK & 86.00\% & 92.11\% & 96.00\% & 98.51\%
\tabularnewline
\hline
EPF & 89.37\% & 93.34\% & 97.42\% & 98.90\%
\tabularnewline
\hline
SC-MK & \textbf{97.21\%} & 97.83\% & 98.11\% & 98.42\%
\tabularnewline
\hline
MFASR  & 95.67\% & 97.88\% & 98.82\% & 99.25\%
\tabularnewline
\hline
2-stage & 96.98\% & \textbf{98.83\%} & \textbf{99.61\%} & \textbf{99.93\%} \tabularnewline \hline
Difference from the best & 0.23 \% & 0.00 \% & 0.00 \% & 0.00\%
\tabularnewline \hline
\end{tabular}
\end{table}

\begin{table}[h!]
\centering
%\begin{center}
\caption{Classification results on the University of Pavia data with different levels of training pixels.\label{trainper_paviaU}}
\begin{tabular}{|c|c|c|c|c|c|}
\hline
Method \textbackslash Training percentage & 4\% & 8\% & 16\% & 32\% \tabularnewline
\hline
$\nu$-SVC & 89.16\% & 91.19\% & 94.04\% & 94.63\%
\tabularnewline
\hline
SVM-CK & 96.80\% &97.93 \% & 98.78\% & 99.13\%
\tabularnewline
\hline
EPF & 97.60\% & 98.37\% & 98.60\% & 98.94\%
\tabularnewline
\hline
SC-MK & 98.83\% & \textbf{99.67\%} & 99.66\% & 99.86\%
\tabularnewline
\hline
MFASR & \textbf{99.02\%} & 99.52\% & 99.81\% & 99.85\%
\tabularnewline
\hline
2-stage & 98.89\% & 99.58\% & \textbf{99.82\%} & \textbf{99.89\%} \tabularnewline \hline
Difference from the best  & 0.13 \% &  0.09\% & 0.00 \% & 0.00\%
\tabularnewline \hline
\end{tabular}
\end{table}

\begin{table}[h!]
\centering
%\begin{center}
\caption{Classification results on the Pavia Center data with different levels of training pixels.\label{trainper_paviacenter}}
\begin{tabular}{|c|c|c|c|c|}
\hline
Method \textbackslash Training percentage &  1\%& 2\% & 4\% & 8\% \tabularnewline
\hline
$\nu$-SVC  & 97.54\% & 98.01\% & 98.28\% & 98.51\%
\tabularnewline
\hline
SVM-CK & 98.80\% & 99.46\% & 99.67\% & 99.83\%
\tabularnewline
\hline
EPF  & \textbf{99.59\%} & \textbf{99.76\%} & 99.76\% & 99.92\%
\tabularnewline
\hline
SC-MK & 99.31\% & 99.59\% & 99.75\% & 99.85\%
\tabularnewline
\hline
MFASR & 99.33\% & 99.64\% & 99.86\% & 99.92\%
\tabularnewline
\hline
2-stage & 99.42\% & 99.73\% & \textbf{99.90\%} & \textbf{99.94\%} \tabularnewline \hline
Difference from the best & 0.17 \%&  0.03\% & 0.00 \% & 0.00\%
\tabularnewline \hline
\end{tabular}
\end{table}

\subsubsection{Model Complexity and Computational Time}
\Cref{parameter_table,runtime_table} shows the computational time required and the number of parameters for all methods. We note that the reported timing does not count the time required to find the optimal set of parameters.
The $\nu$-SVC, SVM-CK and EPF methods have fast computational time because of the simpleness of their models. They have only a few parameters (2, 3 and 4 respectively). However, from the results in \Cref{sec:comparison_results}, they are worse than the other three methods. The SC-MK method is a good method in terms of accuracy and timing, but it has 9 parameters. The MFASR method has 10 parameters and the longest computational time.
In comparison, our method has 5 parameters (2 parameters $\nu$ and $\sigma$ for the $\nu$-SVC (\ref{nu_svm}) and the RBF kernel (\ref{RBFkernel}) respectively in the first stage, 2 parameters $\beta_1$ and $\beta_2$ for the denoising model (\ref{eqn:l2l1anisoll2_eq}) in the second stage and 1 parameter $\mu$ for the ADMM algorithm (\ref{Largrangian})). It has much better (if not the best) classification accuracies and slightly longer computational time than those of $\nu$-SVC, SVM-CK and EPF.

\begin{table}[h!]
\begin{centering}
\caption{Comparison of number of parameters.\label{parameter_table}}
\begin{tabular}{|c c|c|c|c|c|c|c|}
\hline
& & $\nu$-SVC & SVM-CK & EPF &SC-MK& MFASR & 2-stage 
\tabularnewline
\hline
  & Number of parameters&2&3&4&9&10&5
  \tabularnewline
\hline
\end{tabular}

\par\end{centering}
\end{table}

\begin{table}[h!]
\begin{centering}
\caption{Comparison of computational time (in seconds)\label{runtime_table}}
\begin{tabular}{|c c|c|c|c|c|c|c|}
\hline
Data & size/training \% & $\nu$-SVC & SVM-CK & EPF &SC-MK& MFASR & 2-stage
  \tabularnewline
\hline
  Indian Pines & $145\times 145\times 200 /10\%$ & 5.98 & 6.32 & 6.92 & 9.44 & 119 & 8.24
  \tabularnewline
\hline
University of Pavia & $610\times 340 \times 103 /4\%$ & 24.02 & 32.12 & 28.53 & 39.47 & 443 & 35.97
  \tabularnewline
\hline
Pavia Center & $1096\times 715 \times 102 /1\%$ & 58.46 & 81.63 & 118 & 107 & 2599 & 145
  \tabularnewline
\hline
\end{tabular}

\par\end{centering}
\end{table}

\subsection{Effect of the Second-order Term}\label{sec:effect_higher_order}
\begin{figure}[h!]
 \centering
        \begin{subfigure}[t]{0.23\textwidth}
            \centering
            \includegraphics[width=\textwidth]{figures/z_IndianPines_gt_gray.pdf}
            \caption[]%
            {{\small Ground Truth}}
        \end{subfigure}
        \hfill
        \begin{subfigure}[t]{0.23\textwidth}
            \centering
            \includegraphics[width=\textwidth]{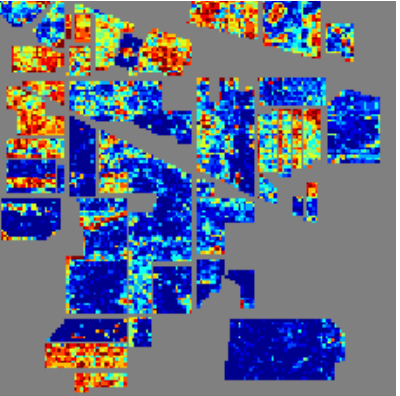}
            \caption[]%
            {{\small $\nu$-SVC \label{SVM_noisy_heatmap}}}
        \end{subfigure}
        \hfill
        \begin{subfigure}[t]{0.23\textwidth}
            \centering
            \includegraphics[width=\textwidth]{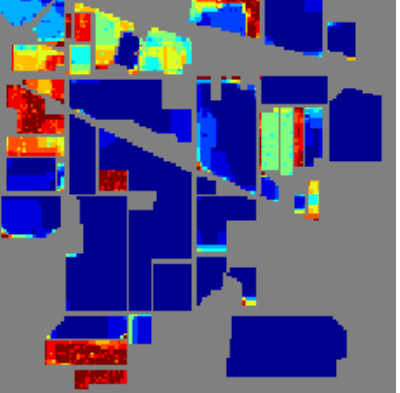}
            \caption[]%
            {{\small 2-stage without $||\nabla \mathbf{u}||_2^2$}}
        \end{subfigure}
        \hfill
        \begin{subfigure}[t]{0.23\textwidth}
            \centering
            \includegraphics[width=\textwidth]{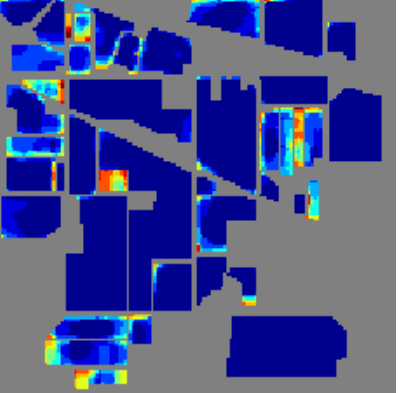}
            \caption[]%
            {{\small 2-stage with $||\nabla \mathbf{u}||_2^2$}}
        \end{subfigure}
        \caption{Heatmaps of mis-classifications on Indian Pines data. (a) ground-truth labels, (b) $\nu$-SVC (the first stage), (c) and (d) our method without or with the second order term respectively. \label{comparison_TV2}}
    \end{figure}

Here we examine empirically the importance of the term $||\nabla \mathbf{u}||_2^2$ in (\ref{eqn:l2l1anisoll2_eq}).
\Cref{comparison_TV2} shows the heatmaps of mis-classifications on the Indian Pines data by
using our method with and without $||\nabla \mathbf{u}||_2^2$ over ten runs. The training pixels are randomly selected and consist of 2.5\% of the labeled pixels.
\Cref{comparison_TV2} (a) shows the ground-truth labels. \Cref{comparison_TV2} (b)--(d) show the heatmaps of mis-classifications of the $\nu$-SVC classifier (i.e. the first stage of our method), the second stage of our method without the $||\nabla \mathbf{u}||_2^2$ term, and the second
stage of our method with the $||\nabla \mathbf{u}||_2^2$ term respectively. Recall the term $||\nabla \mathbf{u}||_2^2$  control  the smoothness of the restored votes and the final classification result is determined by taking the maximum over the restored votes of each class. By choosing the parameter associated with the term appropriately, we can then control the level of shrinking or expanding the homogeneous regions in the final classification result. From \Cref{comparison_TV2} (c), when the term is dropped, the mis-classification regions at the top left and bottom left of the first stage result are not only still mis-classified, but the numbers of mis-classification increase. In contrast, when the term is kept, we see from \Cref{comparison_TV2} (d) that the numbers of mis-classification are significantly lowered. Moreover, most of the mis-classified regions of the first stage result are now correctly classified when the parameters are chosen appropriately.

\section{Conclusions}\label{sec:conclusion}
In this paper, a novel two-stage hyperspectral classification method inspired by image denoising is proposed. The method is simple yet performs effectively. In the first stage, a support vector machine method is used to estimate the pixel-wise probability map of each class. The result in the first stage has decent accuracy but is noisy. In the second stage, an image denoising method is used to clean the probability maps. Since both spectral and spatial information are effectively utilized, our method is very competitive when compared with state-of-the-art classification methods. It also has a simpler framework with fewer number of parameters and faster computational time. It performs particularly well when the inter-class spectra are close or when the training percentage is high.

For future work, we plan to investigate automated parameter selection \cite{Liao2009,Dong2011,Wen2012,Bredies2013} of the denoising method in the second stage, using deep learning methods in the first stage \cite{Yue2015,Makantasis2015,Morchhale2016,Pan2017} and classifying fused hyperspectral and LiDAR data \cite{Gader2013,Debes2014}.
\clearpage
\section*{Acknowledgement}
The authors would like to thank Computational Intelligence Group from the Basque University for sharing the hyperspectral data sets in their website\footnote{\url{http://www.ehu.eus/ccwintco/index.php/Hyperspectral_Remote_Sensing_Scenes}}, Prof. Leyuan Fang from College of Electrical and Information Engineering at Hunan University for providing the programs of the SC-MK and MFASR methods in his homepage\footnote{\url{http://www.escience.cn/people/LeyuanFang}} and Prof. Xudong Kang from College of Electrical and Information Engineering at Hunan University for providing the program of the EPF method in his homepage\footnote{\url{http://xudongkang.weebly.com/}}.

\clearpage

\bibliographystyle{ieeetr}

%\bibliography{2stage_classification_arXiv}
\end{document}